\documentclass[aps,reprint,superscriptaddress]{revtex4-2}
\usepackage{graphicx}
\usepackage{xcolor}

\begin{document}
	
	
	\title{Low Temperature Properties of Low-Loss Macroscopic Lithium Niobate Bulk Acoustic Wave Resonators} 
	
	
	
	\author{William M. Campbell}
	\email{william.campbell@uwa.edu.au}
	\affiliation{Quantum Technologies and Dark Matter Labs, Department of Physics, University of Western Australia, 35 Stirling Highway, Crawley, WA 6009, Australia.}
	\author{Sonali Parashar}\affiliation{Quantum Technologies and Dark Matter Labs, Department of Physics, University of Western Australia, 35 Stirling Highway, Crawley, WA 6009, Australia.}
	\author{Leonardo Mariani}\affiliation{Department of Physics "Giuseppe Occhialini",  University of Milano-Bicocca, Via Festa del Perdono, 7, 20122 Milano MI, Italy}
	\author{Michael E. Tobar}\affiliation{Quantum Technologies and Dark Matter Labs, Department of Physics, University of Western Australia, 35 Stirling Highway, Crawley, WA 6009, Australia.}
	\author{Maxim Goryachev}\affiliation{Quantum Technologies and Dark Matter Labs, Department of Physics, University of Western Australia, 35 Stirling Highway, Crawley, WA 6009, Australia.}
	
	
	\date{\today}
	
	\begin{abstract}
		We investigate gram scale macroscopic bulk acoustic wave (BAW) resonators manufactured from plates of piezoelectric lithium niobate. The intrinsic competing loss mechanisms were studied at cryogenic temperature through precision measurements of various BAW modes.  Exceptional quality factors were measured for the longitudinal BAW modes in the 1-100 MHz range, with a maximum quality factor of 8.9 million, corresponding to a quality factor $\times$ frequency product of 3.8 $\times 10^{14}$ Hz. Through measurements of the acoustic response to a strong drive tone, anomalous self induced absorption and transparency effects are observed. We show that such observations can be explained by microscopic impurities and defect sites in the crystal bulk by the use of a non linear model of acoustic dissipation. The losses associated with these defects provide the ultimate limit of resonator performance, which could be improved in the future if more pure samples were available.
	\end{abstract}
	
	\maketitle 
	\section{Introduction}
		\begin{figure}
		\centering
		\includegraphics[width=0.4\textwidth]{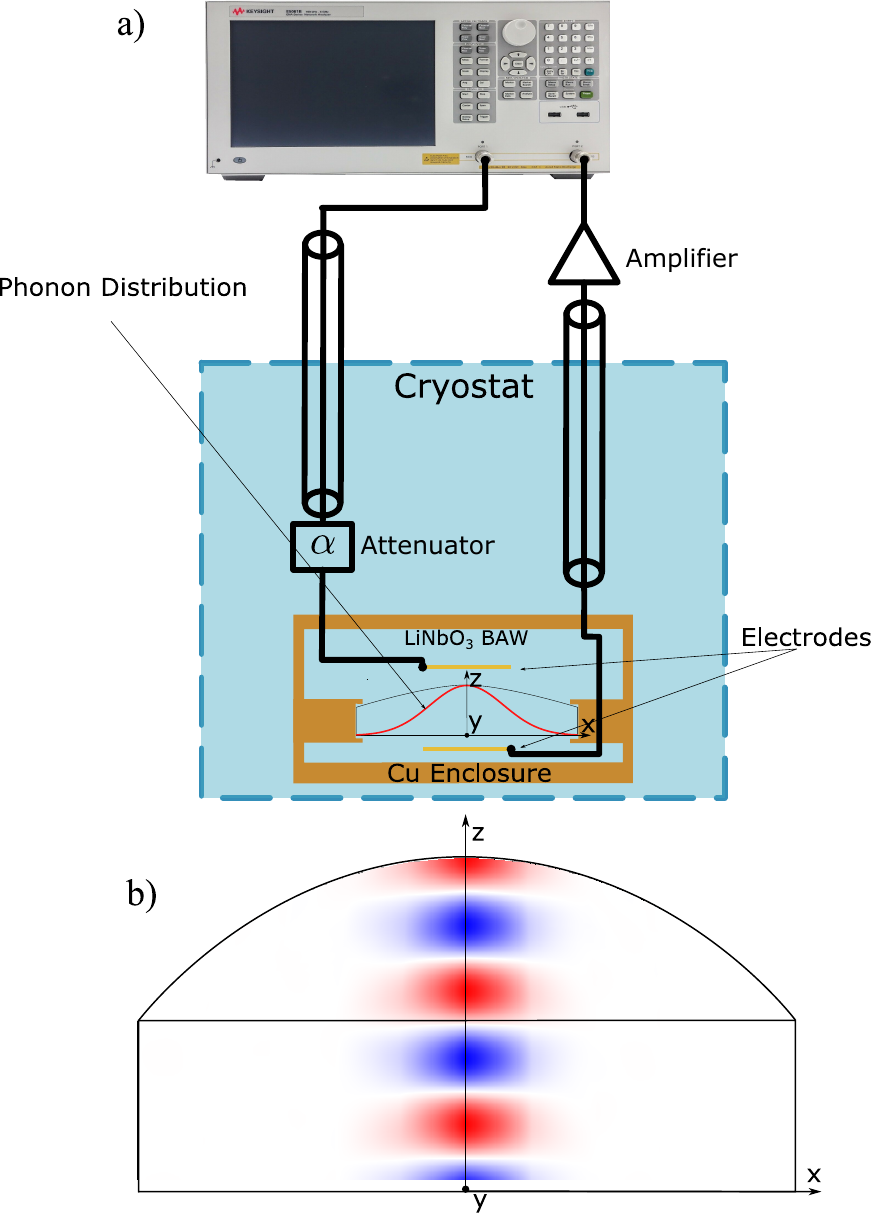}
		\caption{\label{fig:LiNbO3 setup}a) Diagram of experimental set up. Cryostat environment was held under vacuum at temperatures of either 300 K or 4 K, and 10mK depending on the required measurement. The signal path includes attenuation $\alpha$ on the input line in order to reduce thermal noise on the drive signal, as well as an amplifier on the return line to maximise signal to noise ratio. b) Simulated mode distribution of the fifth order longitudinal mode. Red (blue) denotes the positive (negative) displacement in the $z$ direction}
	\end{figure}
	Bulk acoustic wave (BAW) resonators are crystalline devices engineered to sustain long-lived elastic vibrations propagating along the wave vector aligned with the crystal bulk. Historically, these devices have primarily been used for frequency standards and timing applications due to their inherent frequency stability and low acoustic loss \cite{Liu2020}. Beyond engineering applications, the capability to support low-loss phonons makes them desirable for various tests of fundamental physics. For example, BAW phonons can be utilised for quantum information storage and computation in the field of quantum acoustic dynamics \cite{Gokhale2020, Chu2017,Kharel2018, Chu2018}, tests of quantum mechanical principles on large scales \cite{PhysRevLett.130.133604, Bushev2019}, platforms for sensitive optomechanics \cite{Kharel2019}  as well as sensors and detectors for beyond standard model physics, including searches for dark matter \cite{Arvanitaki2016,Carney_2021}, high frequency gravitational waves \cite{rareDetections,Campbell2021, Campbell2023mage, Campbell2023,Aggarwal2021} and tests of Lorentz invariance \cite{Goryachev2018,Lo2016}.

	Low-loss BAW resonators are commonly manufactured from high purity crystalline quartz. This is partly owing to the fact that quartz displays significant improvements to acoustic loss when cooled to cryogenic temperatures\cite{ScRep, Goryachev:2013ly, apl1, apl2}, making such devices ideal for low-energy precision sensing applications, such as tests of fundamental physics. As the requirements for such physics tests become more stringent however, it is natural to further the technological platform by considering alternative piezoelectric materials with differing properties.

	Piezoelectric lithium niobate for example provides the vibrating media for tunable micro-ring resonators \cite{Guarino2007, Zhang:17},  optical frequency combs \cite{Zhang2019}, micro electromechanical system (MEMS) based resonators \cite{Gong2013},  devices for optical to microwave transduction \cite{Marink2021, Shen2020, Zorzetti2023, Wang2022, 10.1063/5.0233800} and acoustic quantum state operations \cite{Wollack2021, Kitzman2023, Kitzman2023b}. This material presents an attractive candidate as a macroscopic BAW resonator due to a stronger piezoelectric coupling coefficient, and smaller loss tangent than crystalline quartz at room temperature. However, material properties display a strong temperature dependence, and are thus drastically different when cooled. This motivates further investigation into the performance of lithium niobate BAWs at cryogenic temperatures, for the purpose of next generation tests of fundamental physics. Additionally, the study of BAW modes at cryogenic temperatures allows one to investigate fundamental properties of the material.

	In this work the loss mechanisms of lithium niobate BAW resonators are investigated by precise measurement of the quality factor ($Q$) of their acoustic modes at cryogenic temperatures. In the following sections \ref{sec:set up} and \ref{sec:nodes}, the measurement set up is introduced and the frequencies of various fundamental bulk vibration modes are identified. Section \ref{sec:loss model} then derives a theoretical model for the total $Q$ of a BAW resonator subject to accumulative loss mechanisms. In sections \ref{sec:Losses} and \ref{sec:temp} the measured $Q$ of various resonator modes are presented as functions of frequency and temperature respectively. Multiple competing loss mechanisms are identified based upon these observations. In section \ref{sec:NL}, we report on a strong drive amplitude dependence on the measured losses. Such self induced absorption and transparency effects are confirmed to be due to microscopic impurities by the use of an appropriate model of non-linear dissipation. Lastly, section \ref{sec:10mK} discusses the performance of these devices when further cooled to the extreme sub-Kelvin temperature of 10 mK.\\
	\\
	\\

	\section{\label{sec:set up} Measurement Set Up}
		\begin{figure*}
		\centering
		\includegraphics[width=0.8\textwidth]{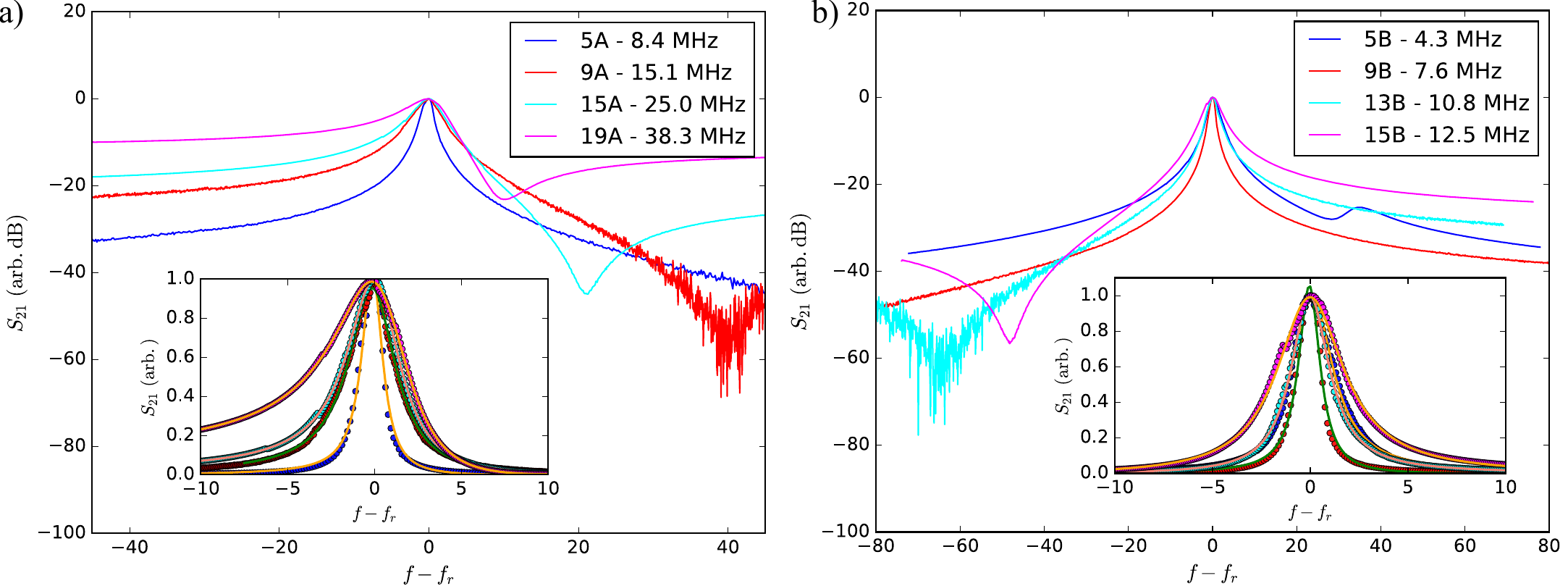}
		\caption{\label{fig:fits}Transmission spectra of A and B polarisation modes are shown in figures a) and b) respectively, plotted for various modes in arbitrary logarithmic units normalised to the resonant peak. Insets in each figure show the same spectra in linear units with the corresponding fit that is used to extract linewidth $\gamma$.}
	\end{figure*}
	The resonator in question consist of a 2 mm thick Z-cut lithium niobate disc, 30 mm in diameter. This geometry was chosen to emulate the typical standard sizes of commercially available macroscopic quartz resonators. One surface is machined with a plano-convex finish specified by a radius of curvature of 100 mm in order to minimise phonon losses via anchoring of the crystal at its edges \cite{Goryachev:2014aa}. The crystal plate is placed inside a copper cavity where it is clamped along its outer edge, with two electrodes placed $\approx$ 1 mm from the crystal surface. Each electrode can then be connected to co-axial lines that run out of the cryostat. The line that carries the excitation signal incident upon the resonator contains -20 dB of attenuation at the cryostat plate. This ensures that 300 K thermal noise on the drive signal is suppressed when operating at 4 K. An amplifier is also placed on the output line such that the returning signal is raised further above the instrument noise floor.
		\begin{figure}
		\includegraphics[width=0.5\textwidth]{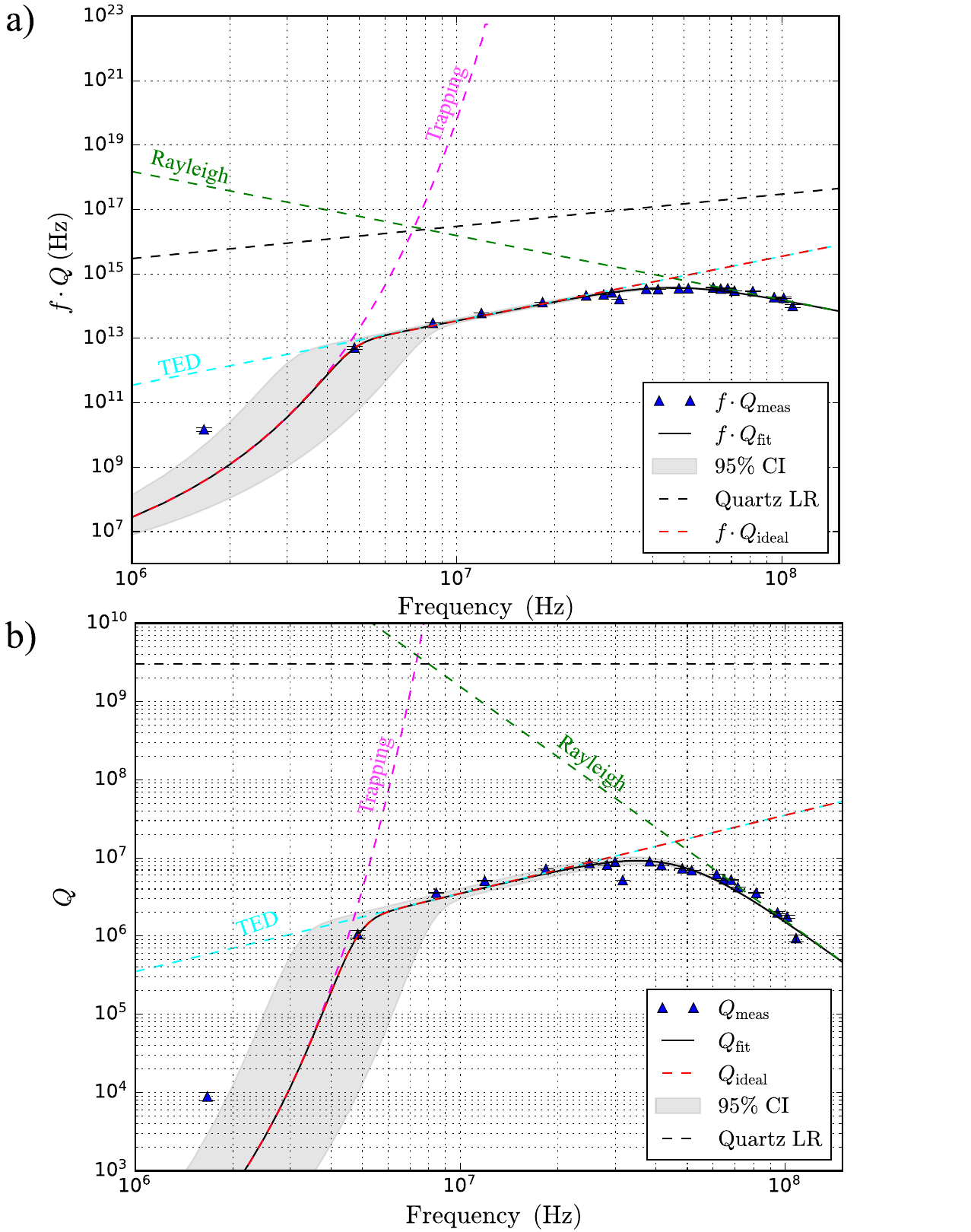}
		\centering
		\caption{\label{fig:LiNbO3-Qplot}Measurement of a)$f \cdot Q$ product and b) intrinsic $Q$ as a function of frequency for various longitudinal modes in the crystal resonators at $T=4$ K. The measured intrinsic quality factors are represented in blue, the black trace and grey bands denote the best fit and 95\% confidence interval to the model of equation (\ref{eq:Qtotal}). The individual influence of clamping, thermoelastic damping (TED) and Rayleigh scattering are shown by the dashed magneta, cyan and green traces respectively. The red trace shows the idealistic performance attainable without the presence of Rayleigh scattering. And the dashed black line gives the intrinsic Landau Rumer limit to losses observed in similar quartz devices \cite{quartzPRL}.}
	\end{figure}
	In this configuration, as shown in figure \ref{fig:LiNbO3 setup}, acoustic resonances can be excited by an electrical drive signal applied to the electrodes. By connecting each co-axial line to ports of a vector network analyser; the acoustic mode frequencies $f_r$ are identified as peaks of maximum power transmission $S_{21}$. Examples of such spectra are presented in figure \ref{fig:fits}. The input drive signal was varied across the frequency spectrum with 1 Hz resolution such that narrow resonances can be identified. The resulting spectra are then compared to a finite element model such that the mode structure of strongly coupled resonances can be identified.

	\section{\label{sec:nodes}Mode Structure}
	Two polarisation families of acoustic modes can be observed in the crystal due to the trigonal symmetry axis of lithium niobate. Longitudinally polarised resonances with bulk displacement in the crystal Z direction are denoted as A modes, while shear wave polarised modes with dominant displacement in the x-y plane are denoted B. In an ideal system only modes with odd overtone number $n$ can be coupled to external electrical signals as even $n$ generates a symmetric electric potential across the crystal. However, in the samples measured, even order modes can be observed with weak coupling. This is ascribed to crystal imperfections and asymmetries allowing for slightly asymmetric electric potentials in the vicinity of $n = \mathrm{even}$ resonances.
	
	The frequencies of strongly coupled modes found by measurement were matched to the frequencies given by solutions to a finite element modelling as in figure \ref{fig:LiNbO3 setup} b), the results of which are presented in table \ref{tab:LiNbO3 modes}. Close agreement between experimental data and model solutions is found. All odd overtones from $n=3$, to $n=15$, in both A and B polarisation are recorded in the table. For higher frequencies, the finite element resolution of the model needs to be further increased in order to guarantee accuracy. However from the results provided in table \ref{tab:LiNbO3 modes}, a simple linear extrapolation $f_i/n_i = f_j/n_j$ can be utilised to identify any $n_{i}$ overtone, of frequency $f_{i}$, for modes of the same polarisation family given knowledge of another mode.
	
	\begin{table}[ht]
		\begin{tabular}{llllllll}
			\hline
			\hline
			$X_{n,m,p}$ & $f_r^\mathrm{(model)}$ & $f_r^\mathrm{(exp)}$ &  $X_{n,m,p}$ & $f_r^\mathrm{(model)}$ & $f_r^\mathrm{(exp)}$\\
			\hline
			\hline
			B$_{5,0,0}$ & 4.08 & 4.20 & A$_{7,0,0}$ & 11.63 & 11.78 \\
			A$_{3,0,0}$ & 4.90 & 4.75 & B$_{15,0,0}$ & 12.22 & 12.24 \\
			B$_{7,0,0}$ & 5.70 & 5.793 & A$_{9,0,0}$ & 14.94 & 14.98\\
			B$_{9,0,0}$ & 7.34 & 7.40 &  A$_{11,0,0}$ & 18.279 & 18.26\\
			A$_{5,0,0}$ & 8.31 & 8.33 &  A$_{13,0,0}$ & 21.60 & 21.48\\
			B$_{11,0,0}$ & 8.96 & 9.02 &  A$_{15,0,0}$ & 24.925 & 24.8\\
			B$_{13,0,0}$ & 10.60 & 10.64 & - & - & - \\
			\hline
			\hline
		\end{tabular}
		\caption{\label{tab:LiNbO3 modes} Frequencies in MHz of the fundamental modes found via finite element modelling $f_r^\mathrm{(model)}$, compared to the experimentally found frequencies $f_r^\mathrm{(exp)}$.}
	\end{table}
	\section{\label{sec:loss model}Theoretical Model of Acoustic Loss}
	The dominant loss mechanisms in BAW resonators can be understood by measurement of the intrinsic quality factor of the acoustic modes $Q$. This quantity is the inverse of the total phonon losses and as such can be parameterised as the inverse sum of the quality factor associated with each competing loss mechanism. For bulk waves, the contributing losses to $Q$ can be due to clamping of the vibrational structure to external supports, thermoelastic damping, phonon-phonon scattering, and phonon-defect scattering. Phonon-electron scattering loss can be dominant in semi-conductor systems or at microwave frequencies, however it is negligible here.
	
	For acoustic resonators at 300 K, phonon-phonon scattering loss is well described by the macroscopic Akheiser regime \cite{Akheiser}, which predicts a constant $Q\cdot f$ product. For lithium niobate the theoretical limit to $Q\cdot f$ as predicted by the Akheiser model is $Q\cdot f \approx 4.74\times10^{14}$, although the largest room temperature value observed in the devices measured in this work is $1.12\times10^{14}$ corresponding to a maximum $Q$ of $3.28\times10^{6}$ . However, at lower temperatures the life time of thermal phonons $\tau_\mathrm{th}$ increases such that phonon attenuation process undergoes a transition into the microscopic description of Landau and Rumer \cite{landaurumer1}. In this regime, the attenuation displays a weaker frequency dependence. As a result  $Q\cdot f$ begins to scale as  $Q\cdot f \propto f$, and $Q$ also greatly increases with decreasing temperature. It is precisely this change to the fundamental limit of intrinsic acoustic loss that gives rise to the numerous applications of high overtones BAW devices at low temperature.  For lithium niobate at 4 K phonon modes are well described by the Landau-Rumer model for $f\gtrsim 1.4$ MHz, which equivalently gives $2\pi f \tau_\mathrm{th} > 1$.
	
	Thermoelastic damping occurs in longitudinal bulk vibrations as the elastic deformation of the travelling wave couples to the thermal expansion tensor associated with local temperature perturbations. Thus longitudinal modes can dissipate energy through the materials thermal conductivity. In the regime where the acoustic wave propagation can be considered quasi-isothermal, the quality factor associated with thermoelastic damping may be written as \cite{ScRep, Deresiewicz1957}
	\begin{equation}\label{eq:TED}
		Q_\mathrm{TED} = \frac{2\pi f}{f^*}\frac{\rho C^s c_{33}}{\lambda_{33}^2 T} ~~~\lambda_{33} := 2c_{12}\alpha_{11} + c_{33}\alpha_{33},
	\end{equation} 
	where $\rho$ is the material density, $T$ is the reference temperature, $C^s$ is the specific heat capacity at constant strain, $c_{ij}$ is the isothermal elastic tensor and $\alpha_{ij}$ the thermal expansion tensor. The turning point frequency $f^*$ denotes the transition point between the adiabatic and isothermal propagation regimes. It is given by $f^*=V_3^2/2\pi\chi_3$ where $\chi_3$ is the material diffusivity. For lithium niobate at $T=4$ K the thermal conductivity and specific heat have been measured, but the elastic constants and expansion coefficients are unknown. Approximating the elastic constants from their room temperature values gives $f^*\approx 13$ MHz, with $f>f^*$ defining the isothermal regime.
	
	For a convex curved vibrating plate, phonons are effectively trapped and confined to the centre, suppressing phonon escape through the support structure and thus limiting the clamping losses. However for low order modes this attenuation is still significant, and has been previously approximated \cite{Goryachev:2014aa} by considering the ratio of vibrational energy in a finite curved plate compared to the total energy of an infinite plate:
	\begin{equation}\label{eq:clamping}
		Q_\mathrm{Clamping} = \left(1-\mathrm{erf}\left(\eta\sqrt{n}\right)^2\right)^{-1}.
	\end{equation}
	Where $n$ is the mode overtone number in the thickness direction and $\eta$ is a dimensionless trapping parameter \cite{Goryachev:2014ab} that should be greater than unity for well trapped phonons. It is clear from this equation that the clamping loss rapidly decreased with increasing $n$.
	
	Acoustic attenuation due to either structural defects or microscopic impurities can be harder to quantify exactly, however scaling relationships can be predicted in some cases. Considering the standard tunnelling model of two level systems in amorphous glasses \cite{Anderson1972, Phillips1972}, it can be shown \cite{Seoanez2007} that the acoustic absorption by an ensemble of low lying two level systems states, potentially arising from non-crystalline impurities, results in a temperature dependent scaling of the acoustic losses $1/Q_\mathrm{TLS}\propto T^{1/3}$. At lower temperatures populations of these two level systems can become unsaturated and have a dominant impact on total acoustic dissipation.
	
	In addition to low lying two level system energy states, Rayleigh-type scattering of phonons in metals and glasses off large impurity ions is predicted to cause attenuation that scales with acoustic wavelength $1/Q_\mathrm{Rayleigh} \propto f^{-3}$ \cite{Mason1947}. This relationship however, has been observed for high overtones in similar quartz piezoelectric BAW systems \cite{quartzPRL}, suggesting the presence of small populations of impurity ions that cause amorphous behaviour at high frequencies.
	
	Examining the measured intrinsic $Q$'s for various longitudinal overtones in lithium niobate presented in figure \ref{fig:LiNbO3-Qplot}, the presence of multiple competing loss regimes can be identified. An appropriate model for the total losses in the system, as a function of frequency, can be written as
	\begin{equation}\label{eq:Qtotal}
		\frac{1}{Q} = \frac{1}{Q_\mathrm{Clamping}}+\frac{1}{Q_\mathrm{TED}} + \frac{1}{Q_\mathrm{Rayleigh}} + \frac{1}{Q_\mathrm{TLS}} + \frac{1}{Q_\mathrm{Landau}}.
	\end{equation}
	Where $Q_\mathrm{Landau}$ represents the intrinsic limit of phonon-phonon dissipation as per the Landau-Rumer model.
	\section{\label{sec:Losses}Frequency Dependence of Acoustic Loss}
	The intrinsic quality factor of bulk lithium niobate modes can be measured by fitting a Lorentzian model to the $S_{21}$ spectrum of a given mode and extracting its linewidth $\gamma$, such that $Q=f_r/\gamma$. The measurement uncertainty following this procedure is then quantified as the standard error associated to the fit parameters, such fits can be seen in the insets of figure \ref{fig:fits}. As the motional resistance of these modes are far greater than the source impedance of the electrical measurement network, the modes are severely under coupled, and thus the external measured quality factor $Q_e$ corresponds exactly to the intrinsic quality factor of the resonator $Q$. The measured values from this procedure are shown in figure \ref{fig:LiNbO3-Qplot}. The cryogenic quality factors show consistent improvement from room temperature values across all frequencies, with a maximum recorded quality factor of  $8.9\times10^6$.
	
	Assuming the first three terms of equation (\ref{eq:Qtotal}) dominate at $T=4$ K, and substituting equations (\ref{eq:TED}), (\ref{eq:clamping}) and the functional form $Q_\mathrm{Rayleigh} = 10^{a}f^{-3}$ , a fit can be performed to the measured quality factors as a function of frequency with $\lambda_{33}$, $\eta$ and $a$ as fit parameters. The resulting best fit and confidence interval band can also be observed in figure \ref{fig:LiNbO3-Qplot}.
	
	The fit extracts a trapping of $\eta=2.2058 \pm 0.24$ and $\lambda_{33} = 2228.27 \pm 73.67$ Pa K$^{-1}$. Assuming for a moment isotopic thermal expansion such that $\alpha_{12}=\alpha_{33}$, an expansion coefficient of $5.63\times10^{-8}~\mathrm{K}^{-1}$ would be attained. Comparing this value to the known coefficients at room temperature of $\alpha_{12} \approx 15\times10^{-5}~\mathrm{K}^{-1}$ and $\alpha_{33} \approx 4\times10^{-5}~\mathrm{K}^{-1}$ we see that the reduction of temperature reduces thermal expansion by roughly three orders of magnitude.Whilst the exact quantitative nature of this argument relies on a linear extrapolation of the known temperature variation in elastic tensor components at $T=273$ K \cite{Smith1971}, as well as the fact that the there is no reason to believe that thermal expansion is isotropic; the determined vast reduction to the thermal expansion tensor is absolute. Indeed reductions to $\alpha_{ij}$ of similar order have been measured previously in piezoelectric quartz when cooled down to temperatures of 1 K \cite{Barron1982}.
	
	The primary significance of this model is the accuracy in the predicted relationships between $Q$ and $f$ across three different regimes defined by competing dominant loss mechanisms. It is observed that thermoelastic damping limits device performance at cryogenic temperatures for some frequencies of interest, a characteristic that is not observed in quartz \cite{ScRep} due to the larger diffusivity and slower phonon velocities when compared to the measured values for lithium niobate at low temperature \cite{Enciso1998}. Additionally, at high frequencies performance becomes limited by Rayleigh scattering of heavy impurity ions. This impurity loss represents a non-fundamental engineering barrier that could be overcome by better crystal purity, thus an additional idealistic loss model has also been plotted on figure \ref{fig:LiNbO3-Qplot} that shows the theoretical device performance of a pure crystal. Even assuming perfectly pure crystals, the intrinsic loss limit at 4 K in lithium niobate is still orders of magnitude below that of quartz resonators at the same temperature and in the presented frequency range. This is due to the stronger thermoelastic damping in lithium niobate when compared to quartz, which has been observed to follow the theoretical Landau-Rumer dissipation at this temperature and frequency range \cite{quartzPRL}. This observed limit for quartz is shown by the dashed black trace in figure \ref{fig:LiNbO3-Qplot} for direct comparison.
	
	Typical impurity concentrations from the crystal growth procedure record a maximum impurity concentration of fluoride at  $<50$ parts per million, with typical concentrations of most other elements at $\lesssim 1$ parts per million, suggesting that even very small populations of impurities greatly inhibit performance benefits at high frequencies. In order to mitigate Rayleigh scattering loss at these frequencies crystal samples would need ot be produced with impurity populations at the level of parts per billion, as has been seen in nano-mechanical devices \cite{gruenkefreudenstein2025}.
	\section{\label{sec:temp}Temperature Dependence of Acoustic Loss}
			\begin{figure}
		\centering
		\includegraphics[width=0.5\textwidth]{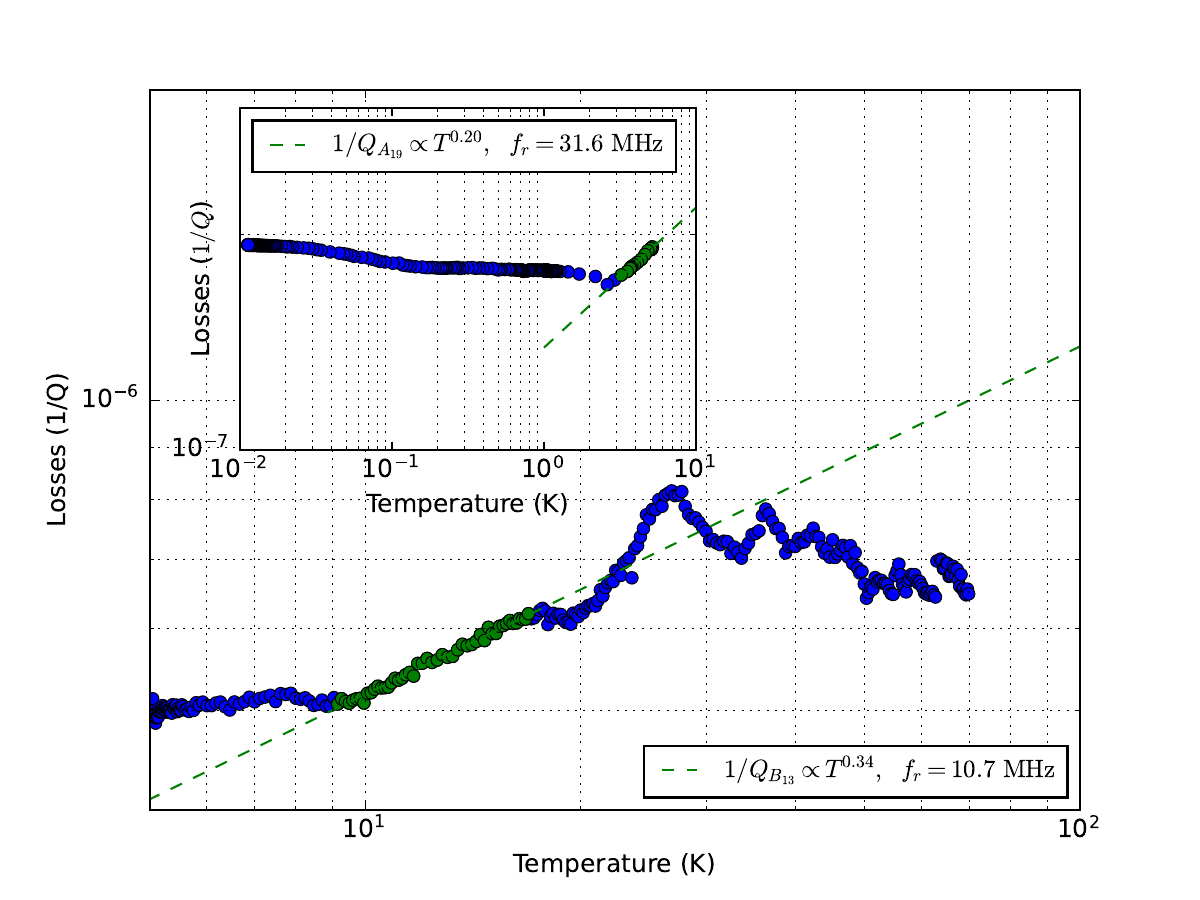}
		\caption{\label{fig:QvT}$Q$ as a function of temperature $T$ is measured for the B$_{13,0,0}$ mode and the resulting scaling relationship between losses $1/Q$ and temperature is shown. Inset shows the scaling for the A$_{19,0,0}$ mode when cooled to 10 mK. Temperature scaling relationships are identified, however of resonant absorption peaks dominate system behaviour.}
	\end{figure}
	The measurement of the quality factor's temperature dependence, $Q(T)$ offers further insights into the dominant loss mechanisms, such as regimes where losses are primarily influenced by an ensemble of low energy two-level systems (TLS) associated with impurity ions. By observing the scaling relationship of losses ($1/Q$) with temperature as shown for the B$_{13,0,0}$ 10.6 MHz mode on the main axes of figure \ref{fig:QvT}, we see the presence of absorption peaks for defined temperature $\approx 28$ K and above. Such increased absorption in narrow temperature regions is due to the system entering an energy regime in which the thermal bath become resonant with phonons. These effects are features of the Landau-Rumer model and have been observed in acoustic systems at low temperature previously. \cite{imbaud:tel-00360494} They are not indicative of TLS induced dissipation.
	
	The characteristic scaling law expected for intrinsic losses due to the coupling of phonons to impurity TLS is theorised to follow $n=1/3$ \cite{Galliou2014, ScRep}, as has been observed in many experimental studies on low temperature acoustic systems \cite{ScRep, TLS1, TLS2, TLS3}. The regime between 10 and 20 K shows a temperature dependence $1/Q \propto T^{0.34}$ that has the correcting scaling to be due to TLS, however the presence of resonant absorption peaks in close proximity suggests that the most dominant loss mechanisms in the $T>10$ K regime are due to thermal phonon effects.
	
	In the inset of figure \ref{fig:QvT}, we show the further cooling of the $A_{19,0,0}$ 31 MHz mode down to $T=$10 mK, where a weaker temperature dependence was observed in the 1-4 K regime, most likely due to the presence of further resonant absorption. At even lower temperatures the losses saturate, showing that the dominating dissipation channel cannot be ascribed to phonon-phonon scattering. However, it is clear that these devices do not reach the regime of TLS loss limited behaviour as the low temperature saturation does not exhibit the appropriate scaling relationship ascribed to TLS dissipation.\\
	\section{\label{sec:NL}Self Induced Transparency and Absorption}
	\begin{figure}
		\includegraphics[width=0.5\textwidth]{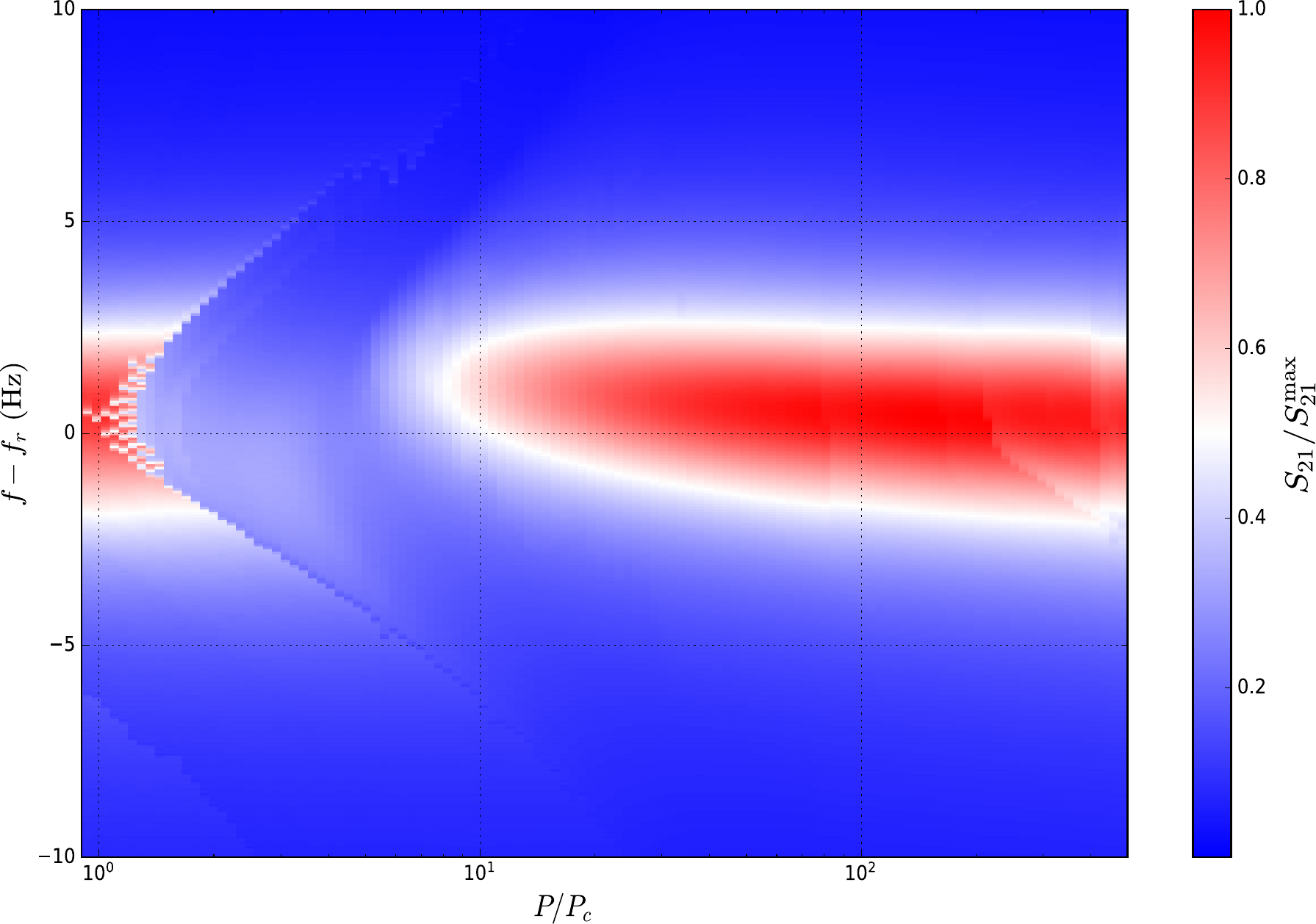}
		\caption{\label{fig:NLexp} Measured $S_{21}$ transmission spectra is shown for the $19^\mathrm{th}$ longitudinal overtone mode as a function of drive power. The drive power is normalised to the critical power value $P_c$, at which point the crystal non-linear regime becomes dominant and the mode begins to demonstrate self induced absorption.}
	\end{figure}
	\begin{figure*}
		\includegraphics[width=\textwidth]{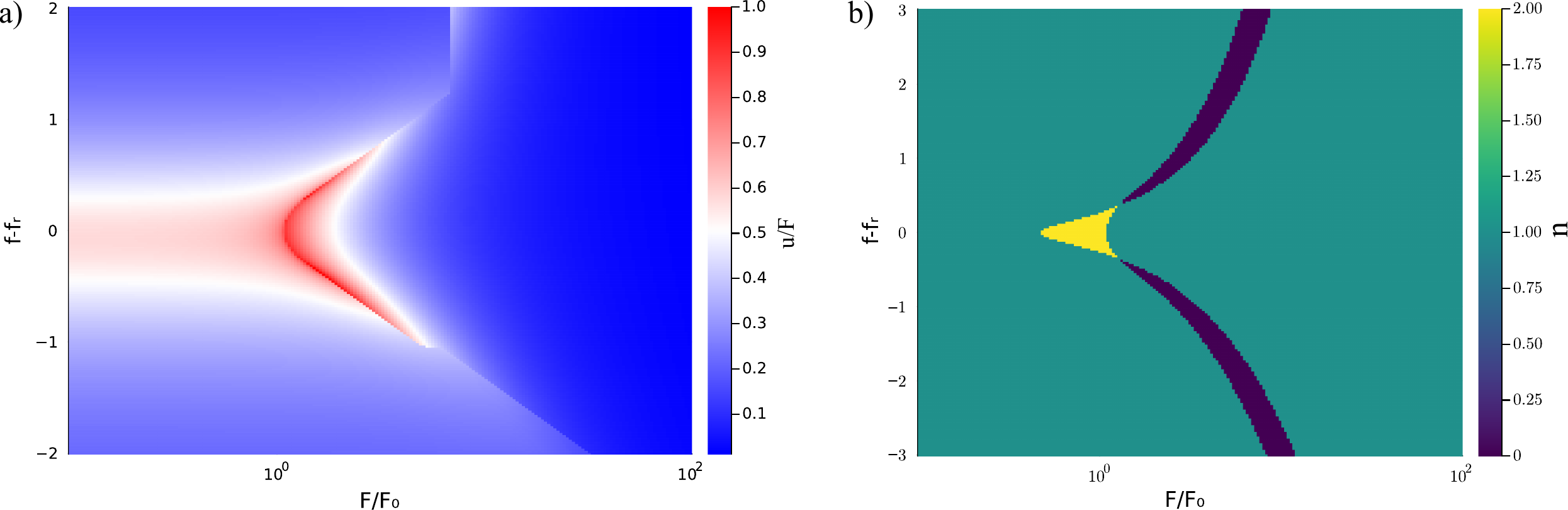}
		\caption{\label{fig:NLsim}Numerical solutions to equation (\ref{eq:NLmodel}) determined via harmonic balance methods are presented. a) The frequency response spectrum of a singular solution branch is plotted as a function of increasing drive force. The drive is normalised by the threshold level that induces absorption $F_0$. b) The parameter space in which stable solutions $n$ exist, is shown. Each of the three stable solution families $n$ are distinguished by a corresponding colour for a given frequency $f$ and force $F$. These simulations demonstrate clear resemblance to the data of figure \ref{fig:NLexp}, in which the bi-stable region is symmetrical about resonance, and gives a drive dependent regime of increased absorption.}
	\end{figure*}
	Investigating the lithium niobate resonators at large drive amplitudes, strongly non-linear regimes are observed. Utilising the same set up as described in section \ref{sec:set up}, the resonator modes transmission spectrum were measured as a function of increasing drive power $P$. An example is presented In figure \ref{fig:NLexp}, where a non linear regime is reached for some critical power level $P_c$ in the $19^\mathrm{th}$ longitudinal overtone mode. After which the transmission collapses, and the frequency response enters a regime of increased dissipation, a phenomena known as self induced absorption. At even larger drive amplitudes, the response ultimately recovers during the inverse process of self induced transparency.
	
	Bulk acoustic wave systems in general are known to enter non-linear regimes in which higher order terms in the vibrational amplitude $u$ begin to dominate system behaviour. A famous example of this is the existence of the Duffing non-linearity present in many acoustic systems \cite{Tiersten1976, Tiersten1986, nonlin}. This effect is characterised by an asymmetrical, amplitude dependent perturbation to the resonant frequency due to high order elastic constants generating terms cubic in $u$. However, the self induced absorption and transparency phenomena of figure \ref{fig:NLexp} cannot be explained by a duffing model, as the discontinuities in frequency response occur symmetrically about resonance.
	
	In so called `un-swept' piezoelectric quartz BAW resonators that are yet to undergo a sweeping impurity removal process, various anomalous nonlinear effects are also observed that cannot be explained by the Duffing model \cite{quartzJAP}. The quartz devices exhibit discontinuities in their frequency response for certain drive amplitudes, van-der-pol like quadrature oscillations, as well as regimes of self induced transparency and absorption. These effects are reported to be due to the existence of heavy impurity ions in the crystalline bulk, prevalent to the levels of parts per million.
	
	Self induced transparency and absorption has also previously been observed in acoustic media with artificially introduced defects \cite{Fillinger2006, FILLINGER2008}. Here the defects were modelled as localised non linear dissipation terms introduced to the resonator's equations of motion. These terms then introduce perturbations on top of the linear losses in certain regimes. In previous work with un-swept quartz analytical arguments were made to show that such a model could qualitatively reproduces the observed phenomena.
	
	Taking the simplified dissipation model for the case of a singular defect \cite{quartzJAP}, and considering a polynomial non-linear dissipation function $f(\dot{u}))$ \cite{Fillinger2006}, we can write the equation of motion
	\begin{equation}\label{eq:NLmodel}
		\ddot{u}+2\gamma(1+f(\dot{u}))\dot{u}+\omega_r^2u= F\mathrm{cos}(\omega t),
	\end{equation}
	where $\omega$ is angular frequency, $\gamma$ parameterises linear dissipation, $F$ is the external driving force to the system, and $f(\dot{u})$ is given by $f(\dot{\alpha}) = -\beta_1 \dot{\alpha}^2+\beta_2\dot{\alpha}^4$ with appropriate choices for the constants $\beta_{1,2}$.
	
	This non linear model can then be solved utilising numerical harmonic balance methods in order to reproduce the anomalous self induced absorption effect. As the system is strongly-nonlinear, a variety of solution branches for $u$ are attained. In figure \ref{fig:NLsim} a) the spectrum of normalised displacement amplitudes $u(f)$ is plotted for one of the solution branches, as a function of drive power $F$. The same symmetrical discontinuity about resonance can be observed as in figure \ref{fig:NLexp}, as the response collapses under self induced absorption due to the presence of non-linear dissipation that dominates above a threshold drive force. By visualising the parameter space for stable solutions to equation (\ref{eq:NLmodel}), plotted in figure \ref{fig:NLsim} b), it is clear that the non-linear dissipation term causes switching between bi-stable states on both sides of the resonant peak. This bi-stability can facilitate the observed van-der-pol like oscillations in physical systems, and also recreates the characteristic self induced absorption phenomena. The general form of equation (\ref{eq:NLmodel}) has been shown previously to admit solutions in which the resonant peak recovers under self induced transparency \cite{quartzJAP}. Alternatively one can utilise a non-diverging non-linear dissipation function $f(\dot{u})$ in order to model self induced transparency.
	
	The accuracy of this model in reproducing the non-linear behaviour observed in figure \ref{fig:NLexp} suggests that small populations of heavy ions limit the performance of lithium niobate BAW resonators, as in un-swept quartz.
	
	\begin{figure}
		\centering
		\includegraphics[width=0.45\textwidth]{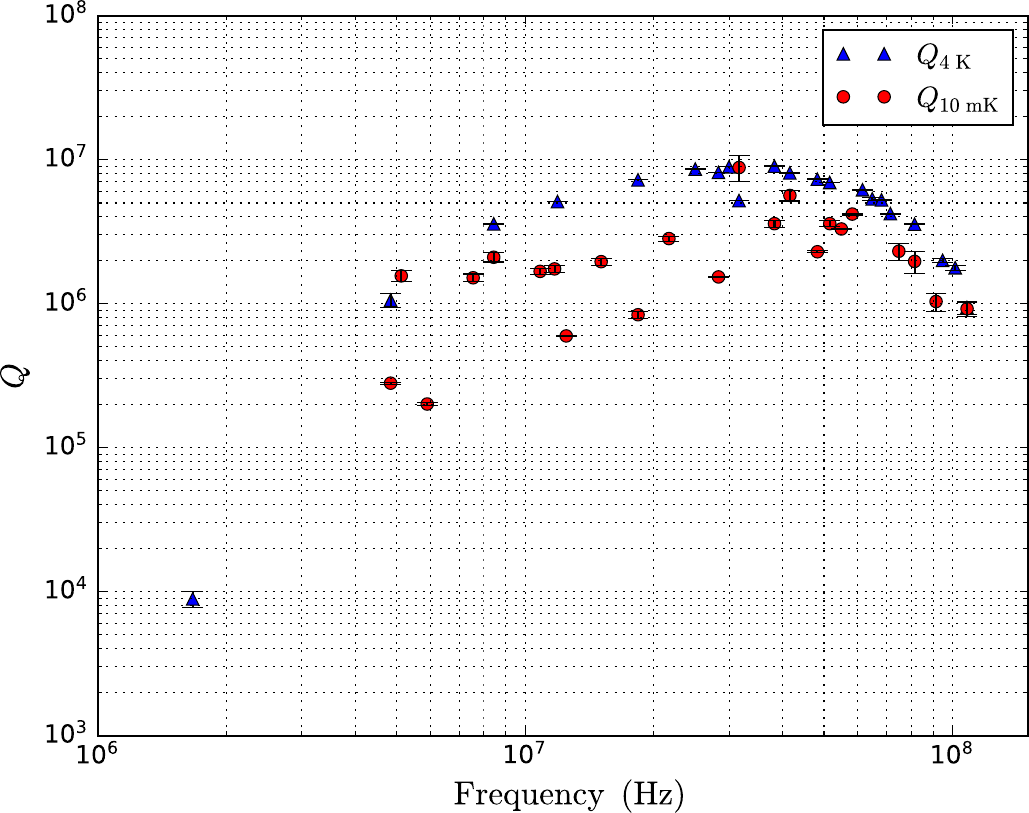}
		\caption{\label{fig:mK} Measured quality factors at $T=4$ K from figure \ref{fig:LiNbO3-Qplot} are plotted with the quality factors for the same modes measured at $T=10$ mK as shown in red. Increased absorption is clearly observed in the sub-Kelvin regime.}
	\end{figure}
	\section{\label{sec:10mK}Performance at Sub-Kelvin Temperature}
	The cryostat system as described in section \ref{sec:set up} can be further cooled from 4 K down to sub-Kelvin temperatures due to helium dilution. This provides an opportunity to explore the lithium niobate system at extreme
	cryogenic temperatures leading to the quantum regime where the thermal occupation number approaches unity. Taking a look at figure \ref{fig:mK} the quality factors greatly suffer when cooled further to 10 mK. Some reduction in $Q$ can occasionally be compensated by driving to larger amplitudes. This feature is reminiscent of the saturation of TLS dissipation, however the effect observed here does not follow well known TLS models \cite{Wollack2021}. Instead, the extra dissipation at this temperature could be described by the oscillator state being deeper into the non-linear regime of section \ref{sec:NL} and thus the saturation comes as the system enters the regime of self-induced transparency. We observe that the critical applied drive power for self-induced absorption to be lower at 10 mK than at 4 K for nearly all modes.

	At sub-Kelvin temperatures a reduction of phonon-phonon losses causes an increase to the intrinsic energy stored in the crystal, and thus the acoustic motion enters a deeper non-linear regime when compared to 4 Kelvin temperatures for the same incident drive power. Such phenomena is well known when it comes to the duffing effect in quartz \cite{quartzJAP, goryachev:hal-00670126}, and could be the cause for the stronger non-linear dissipation observed here. It is important to note that in these strong non-linear regimes the quality factor becomes harder to strictly define with multiple channels contributing to intrinsic dissipation. 
	\section{Conclusion}
	Z-cut lithium niobate provides a promising material for macroscopic BAW resonators designed with the intention of testing areas of fundamental physics, demonstrating large quality factors for bulk acoustic modes. However, the existence of microscopic impurities sites in the crystalline bulk puts limitations on the ultimate performance for high overtones. Additionally, the strong thermoelastic damping of the material results in further attenuation at low frequencies in comparison to crystalline quartz. We demonstrate that the losses in our system are well understood, with intrinsic quality factors adhering to models predicted by theory. Future improvements to the quality factors of modes $f_r>30$ MHz  in lithium niobate resonators can be made by reducing atomic impurity levels. Treatments such as DC sweeping, and improvements to crystal growth procedures have been successfully to produce higher quality material in other systems. Pending such engineering improvements lithium niobate BAW resonators will see high overtone quality factors tend upwards of $10^7$, reaching the intrinsic limit set by thermoelastic damping. Such high quality factors will be beneficial to various fundamental physics experiments hoping to exploit the large mass macroscopic phonon modes provided by this system.
	\section{Acknowledgments}
	This research was supported by the ARC Centre of Excellence for Engineered Quantum Systems (EQUS, No.CE170100009) along with support from the Defence Science and Technology Group (DSTG) as part of the EQUS Quantum Clock Flagship program. Additional support provided by the ARC Centre of Excellence for Dark Matter Particle Physics (CDM, No. CE200100008).
	\bibliography{refs}

\begin{thebibliography}{59}%
\makeatletter
\providecommand \@ifxundefined [1]{%
 \@ifx{#1\undefined}
}%
\providecommand \@ifnum [1]{%
 \ifnum #1\expandafter \@firstoftwo
 \else \expandafter \@secondoftwo
 \fi
}%
\providecommand \@ifx [1]{%
 \ifx #1\expandafter \@firstoftwo
 \else \expandafter \@secondoftwo
 \fi
}%
\providecommand \natexlab [1]{#1}%
\providecommand \enquote  [1]{``#1''}%
\providecommand \bibnamefont  [1]{#1}%
\providecommand \bibfnamefont [1]{#1}%
\providecommand \citenamefont [1]{#1}%
\providecommand \href@noop [0]{\@secondoftwo}%
\providecommand \href [0]{\begingroup \@sanitize@url \@href}%
\providecommand \@href[1]{\@@startlink{#1}\@@href}%
\providecommand \@@href[1]{\endgroup#1\@@endlink}%
\providecommand \@sanitize@url [0]{\catcode `\\12\catcode `\$12\catcode
  `\&12\catcode `\#12\catcode `\^12\catcode `\_12\catcode `\%12\relax}%
\providecommand \@@startlink[1]{}%
\providecommand \@@endlink[0]{}%
\providecommand \url  [0]{\begingroup\@sanitize@url \@url }%
\providecommand \@url [1]{\endgroup\@href {#1}{\urlprefix }}%
\providecommand \urlprefix  [0]{URL }%
\providecommand \Eprint [0]{\href }%
\providecommand \doibase [0]{https://doi.org/}%
\providecommand \selectlanguage [0]{\@gobble}%
\providecommand \bibinfo  [0]{\@secondoftwo}%
\providecommand \bibfield  [0]{\@secondoftwo}%
\providecommand \translation [1]{[#1]}%
\providecommand \BibitemOpen [0]{}%
\providecommand \bibitemStop [0]{}%
\providecommand \bibitemNoStop [0]{.\EOS\space}%
\providecommand \EOS [0]{\spacefactor3000\relax}%
\providecommand \BibitemShut  [1]{\csname bibitem#1\endcsname}%
\let\auto@bib@innerbib\@empty
\bibitem [{\citenamefont {Liu}\ \emph {et~al.}(2020)\citenamefont {Liu},
  \citenamefont {Cai}, \citenamefont {Zhang}, \citenamefont {Tovstopyat},
  \citenamefont {Liu},\ and\ \citenamefont {Sun}}]{Liu2020}%
  \BibitemOpen
  \bibfield  {author} {\bibinfo {author} {\bibfnamefont {Y.}~\bibnamefont
  {Liu}}, \bibinfo {author} {\bibfnamefont {Y.}~\bibnamefont {Cai}}, \bibinfo
  {author} {\bibfnamefont {Y.}~\bibnamefont {Zhang}}, \bibinfo {author}
  {\bibfnamefont {A.}~\bibnamefont {Tovstopyat}}, \bibinfo {author}
  {\bibfnamefont {S.}~\bibnamefont {Liu}},\ and\ \bibinfo {author}
  {\bibfnamefont {C.}~\bibnamefont {Sun}},\ }\bibfield  {title} {\bibinfo
  {title} {Materials, design, and characteristics of bulk acoustic wave
  resonator: A review},\ }\bibfield  {journal} {\bibinfo  {journal}
  {Micromachines}\ }\textbf {\bibinfo {volume} {11}},\ \href
  {https://doi.org/10.3390/mi11070630} {10.3390/mi11070630} (\bibinfo {year}
  {2020})\BibitemShut {NoStop}%
\bibitem [{\citenamefont {Gokhale}\ \emph {et~al.}(2020)\citenamefont
  {Gokhale}, \citenamefont {Downey}, \citenamefont {Katzer}, \citenamefont
  {Nepal}, \citenamefont {Lang}, \citenamefont {Stroud},\ and\ \citenamefont
  {Meyer}}]{Gokhale2020}%
  \BibitemOpen
  \bibfield  {author} {\bibinfo {author} {\bibfnamefont {V.~J.}\ \bibnamefont
  {Gokhale}}, \bibinfo {author} {\bibfnamefont {B.~P.}\ \bibnamefont {Downey}},
  \bibinfo {author} {\bibfnamefont {D.~S.}\ \bibnamefont {Katzer}}, \bibinfo
  {author} {\bibfnamefont {N.}~\bibnamefont {Nepal}}, \bibinfo {author}
  {\bibfnamefont {A.~C.}\ \bibnamefont {Lang}}, \bibinfo {author}
  {\bibfnamefont {R.~M.}\ \bibnamefont {Stroud}},\ and\ \bibinfo {author}
  {\bibfnamefont {D.~J.}\ \bibnamefont {Meyer}},\ }\bibfield  {title} {\bibinfo
  {title} {Epitaxial bulk acoustic wave resonators as highly coherent
  multi-phonon sources for quantum acoustodynamics},\ }\href
  {https://doi.org/10.1038/s41467-020-15472-w} {\bibfield  {journal} {\bibinfo
  {journal} {Nature Communications}\ }\textbf {\bibinfo {volume} {11}},\
  \bibinfo {pages} {2314} (\bibinfo {year} {2020})}\BibitemShut {NoStop}%
\bibitem [{\citenamefont {Chu}\ \emph {et~al.}(2017)\citenamefont {Chu},
  \citenamefont {Kharel}, \citenamefont {Renninger}, \citenamefont {Burkhart},
  \citenamefont {Frunzio}, \citenamefont {Rakich},\ and\ \citenamefont
  {Schoelkopf}}]{Chu2017}%
  \BibitemOpen
  \bibfield  {author} {\bibinfo {author} {\bibfnamefont {Y.}~\bibnamefont
  {Chu}}, \bibinfo {author} {\bibfnamefont {P.}~\bibnamefont {Kharel}},
  \bibinfo {author} {\bibfnamefont {W.~H.}\ \bibnamefont {Renninger}}, \bibinfo
  {author} {\bibfnamefont {L.~D.}\ \bibnamefont {Burkhart}}, \bibinfo {author}
  {\bibfnamefont {L.}~\bibnamefont {Frunzio}}, \bibinfo {author} {\bibfnamefont
  {P.~T.}\ \bibnamefont {Rakich}},\ and\ \bibinfo {author} {\bibfnamefont
  {R.~J.}\ \bibnamefont {Schoelkopf}},\ }\bibfield  {title} {\bibinfo {title}
  {{Quantum acoustics with superconducting qubits}},\ }\href
  {https://doi.org/10.1126/science.aao1511} {\bibfield  {journal} {\bibinfo
  {journal} {Science}\ }\textbf {\bibinfo {volume} {358}},\ \bibinfo {pages}
  {199} (\bibinfo {year} {2017})},\ \Eprint {https://arxiv.org/abs/1703.00342}
  {arXiv:1703.00342} \BibitemShut {NoStop}%
\bibitem [{\citenamefont {Kharel}\ \emph {et~al.}(2018)\citenamefont {Kharel},
  \citenamefont {Chu}, \citenamefont {Power}, \citenamefont {Renninger},
  \citenamefont {Schoelkopf},\ and\ \citenamefont {Rakich}}]{Kharel2018}%
  \BibitemOpen
  \bibfield  {author} {\bibinfo {author} {\bibfnamefont {P.}~\bibnamefont
  {Kharel}}, \bibinfo {author} {\bibfnamefont {Y.}~\bibnamefont {Chu}},
  \bibinfo {author} {\bibfnamefont {M.}~\bibnamefont {Power}}, \bibinfo
  {author} {\bibfnamefont {W.~H.}\ \bibnamefont {Renninger}}, \bibinfo {author}
  {\bibfnamefont {R.~J.}\ \bibnamefont {Schoelkopf}},\ and\ \bibinfo {author}
  {\bibfnamefont {P.~T.}\ \bibnamefont {Rakich}},\ }\bibfield  {title}
  {\bibinfo {title} {Ultra-high-q phononic resonators on-chip at cryogenic
  temperatures},\ }\href {https://doi.org/10.1063/1.5026798} {\bibfield
  {journal} {\bibinfo  {journal} {APL Photonics}\ }\textbf {\bibinfo {volume}
  {3}},\ \bibinfo {pages} {066101} (\bibinfo {year} {2018})},\ \Eprint
  {https://arxiv.org/abs/https://doi.org/10.1063/1.5026798}
  {https://doi.org/10.1063/1.5026798} \BibitemShut {NoStop}%
\bibitem [{\citenamefont {Chu}\ \emph {et~al.}(2018)\citenamefont {Chu},
  \citenamefont {Kharel}, \citenamefont {Yoon}, \citenamefont {Frunzio},
  \citenamefont {Rakich},\ and\ \citenamefont {Schoelkopf}}]{Chu2018}%
  \BibitemOpen
  \bibfield  {author} {\bibinfo {author} {\bibfnamefont {Y.}~\bibnamefont
  {Chu}}, \bibinfo {author} {\bibfnamefont {P.}~\bibnamefont {Kharel}},
  \bibinfo {author} {\bibfnamefont {T.}~\bibnamefont {Yoon}}, \bibinfo {author}
  {\bibfnamefont {L.}~\bibnamefont {Frunzio}}, \bibinfo {author} {\bibfnamefont
  {P.~T.}\ \bibnamefont {Rakich}},\ and\ \bibinfo {author} {\bibfnamefont
  {R.~J.}\ \bibnamefont {Schoelkopf}},\ }\bibfield  {title} {\bibinfo {title}
  {Creation and control of multi-phonon fock states in a bulk acoustic-wave
  resonator},\ }\href {https://doi.org/10.1038/s41586-018-0717-7} {\bibfield
  {journal} {\bibinfo  {journal} {Nature}\ }\textbf {\bibinfo {volume} {563}},\
  \bibinfo {pages} {666} (\bibinfo {year} {2018})}\BibitemShut {NoStop}%
\bibitem [{\citenamefont {Schrinski}\ \emph {et~al.}(2023)\citenamefont
  {Schrinski}, \citenamefont {Yang}, \citenamefont {von L\"upke}, \citenamefont
  {Bild}, \citenamefont {Chu}, \citenamefont {Hornberger}, \citenamefont
  {Nimmrichter},\ and\ \citenamefont {Fadel}}]{PhysRevLett.130.133604}%
  \BibitemOpen
  \bibfield  {author} {\bibinfo {author} {\bibfnamefont {B.}~\bibnamefont
  {Schrinski}}, \bibinfo {author} {\bibfnamefont {Y.}~\bibnamefont {Yang}},
  \bibinfo {author} {\bibfnamefont {U.}~\bibnamefont {von L\"upke}}, \bibinfo
  {author} {\bibfnamefont {M.}~\bibnamefont {Bild}}, \bibinfo {author}
  {\bibfnamefont {Y.}~\bibnamefont {Chu}}, \bibinfo {author} {\bibfnamefont
  {K.}~\bibnamefont {Hornberger}}, \bibinfo {author} {\bibfnamefont
  {S.}~\bibnamefont {Nimmrichter}},\ and\ \bibinfo {author} {\bibfnamefont
  {M.}~\bibnamefont {Fadel}},\ }\bibfield  {title} {\bibinfo {title}
  {Macroscopic quantum test with bulk acoustic wave resonators},\ }\href
  {https://doi.org/10.1103/PhysRevLett.130.133604} {\bibfield  {journal}
  {\bibinfo  {journal} {Phys. Rev. Lett.}\ }\textbf {\bibinfo {volume} {130}},\
  \bibinfo {pages} {133604} (\bibinfo {year} {2023})}\BibitemShut {NoStop}%
\bibitem [{\citenamefont {Bushev}\ \emph {et~al.}(2019)\citenamefont {Bushev},
  \citenamefont {Bourhill}, \citenamefont {Goryachev}, \citenamefont
  {Kukharchyk}, \citenamefont {Ivanov}, \citenamefont {Galliou}, \citenamefont
  {Tobar},\ and\ \citenamefont {Danilishin}}]{Bushev2019}%
  \BibitemOpen
  \bibfield  {author} {\bibinfo {author} {\bibfnamefont {P.~A.}\ \bibnamefont
  {Bushev}}, \bibinfo {author} {\bibfnamefont {J.}~\bibnamefont {Bourhill}},
  \bibinfo {author} {\bibfnamefont {M.}~\bibnamefont {Goryachev}}, \bibinfo
  {author} {\bibfnamefont {N.}~\bibnamefont {Kukharchyk}}, \bibinfo {author}
  {\bibfnamefont {E.}~\bibnamefont {Ivanov}}, \bibinfo {author} {\bibfnamefont
  {S.}~\bibnamefont {Galliou}}, \bibinfo {author} {\bibfnamefont {M.~E.}\
  \bibnamefont {Tobar}},\ and\ \bibinfo {author} {\bibfnamefont
  {S.}~\bibnamefont {Danilishin}},\ }\bibfield  {title} {\bibinfo {title}
  {{Testing the generalized uncertainty principle with macroscopic mechanical
  oscillators and pendulums}},\ }\href
  {https://doi.org/10.1103/PhysRevD.100.066020} {\bibfield  {journal} {\bibinfo
   {journal} {Physical Review D}\ }\textbf {\bibinfo {volume} {100}},\ \bibinfo
  {pages} {66020} (\bibinfo {year} {2019})},\ \Eprint
  {https://arxiv.org/abs/1903.03346} {arXiv:1903.03346} \BibitemShut {NoStop}%
\bibitem [{\citenamefont {Kharel}\ \emph {et~al.}(2019)\citenamefont {Kharel},
  \citenamefont {Harris}, \citenamefont {Kittlaus}, \citenamefont {Renninger},
  \citenamefont {Otterstrom}, \citenamefont {Harris},\ and\ \citenamefont
  {Rakich}}]{Kharel2019}%
  \BibitemOpen
  \bibfield  {author} {\bibinfo {author} {\bibfnamefont {P.}~\bibnamefont
  {Kharel}}, \bibinfo {author} {\bibfnamefont {G.~I.}\ \bibnamefont {Harris}},
  \bibinfo {author} {\bibfnamefont {E.~A.}\ \bibnamefont {Kittlaus}}, \bibinfo
  {author} {\bibfnamefont {W.~H.}\ \bibnamefont {Renninger}}, \bibinfo {author}
  {\bibfnamefont {N.~T.}\ \bibnamefont {Otterstrom}}, \bibinfo {author}
  {\bibfnamefont {J.~G.~E.}\ \bibnamefont {Harris}},\ and\ \bibinfo {author}
  {\bibfnamefont {P.~T.}\ \bibnamefont {Rakich}},\ }\bibfield  {title}
  {\bibinfo {title} {High-frequency cavity optomechanics using bulk acoustic
  phonons},\ }\href {https://doi.org/10.1126/sciadv.aav0582} {\bibfield
  {journal} {\bibinfo  {journal} {Science Advances}\ }\textbf {\bibinfo
  {volume} {5}},\ \bibinfo {pages} {eaav0582} (\bibinfo {year} {2019})},\
  \Eprint
  {https://arxiv.org/abs/https://www.science.org/doi/pdf/10.1126/sciadv.aav0582}
  {https://www.science.org/doi/pdf/10.1126/sciadv.aav0582} \BibitemShut
  {NoStop}%
\bibitem [{\citenamefont {Arvanitaki}\ \emph {et~al.}(2016)\citenamefont
  {Arvanitaki}, \citenamefont {Dimopoulos},\ and\ \citenamefont {{Van
  Tilburg}}}]{Arvanitaki2016}%
  \BibitemOpen
  \bibfield  {author} {\bibinfo {author} {\bibfnamefont {A.}~\bibnamefont
  {Arvanitaki}}, \bibinfo {author} {\bibfnamefont {S.}~\bibnamefont
  {Dimopoulos}},\ and\ \bibinfo {author} {\bibfnamefont {K.}~\bibnamefont {{Van
  Tilburg}}},\ }\bibfield  {title} {\bibinfo {title} {{Sound of Dark Matter:
  Searching for Light Scalars with Resonant-Mass Detectors}},\ }\href
  {https://doi.org/10.1103/PhysRevLett.116.031102} {\bibfield  {journal}
  {\bibinfo  {journal} {Physical Review Letters}\ }\textbf {\bibinfo {volume}
  {116}},\ \bibinfo {pages} {1} (\bibinfo {year} {2016})},\ \Eprint
  {https://arxiv.org/abs/1508.01798} {arXiv:1508.01798} \BibitemShut {NoStop}%
\bibitem [{\citenamefont {Carney}\ \emph {et~al.}(2021)\citenamefont {Carney},
  \citenamefont {Hook}, \citenamefont {Liu}, \citenamefont {Taylor},\ and\
  \citenamefont {Zhao}}]{Carney_2021}%
  \BibitemOpen
  \bibfield  {author} {\bibinfo {author} {\bibfnamefont {D.}~\bibnamefont
  {Carney}}, \bibinfo {author} {\bibfnamefont {A.}~\bibnamefont {Hook}},
  \bibinfo {author} {\bibfnamefont {Z.}~\bibnamefont {Liu}}, \bibinfo {author}
  {\bibfnamefont {J.~M.}\ \bibnamefont {Taylor}},\ and\ \bibinfo {author}
  {\bibfnamefont {Y.}~\bibnamefont {Zhao}},\ }\bibfield  {title} {\bibinfo
  {title} {Ultralight dark matter detection with mechanical quantum sensors},\
  }\href {https://doi.org/10.1088/1367-2630/abd9e7} {\bibfield  {journal}
  {\bibinfo  {journal} {New Journal of Physics}\ }\textbf {\bibinfo {volume}
  {23}},\ \bibinfo {pages} {023041} (\bibinfo {year} {2021})}\BibitemShut
  {NoStop}%
\bibitem [{\citenamefont {Goryachev}\ \emph {et~al.}(2021)\citenamefont
  {Goryachev}, \citenamefont {Campbell}, \citenamefont {Heng}, \citenamefont
  {Galliou}, \citenamefont {Ivanov},\ and\ \citenamefont
  {Tobar}}]{rareDetections}%
  \BibitemOpen
  \bibfield  {author} {\bibinfo {author} {\bibfnamefont {M.}~\bibnamefont
  {Goryachev}}, \bibinfo {author} {\bibfnamefont {W.~M.}\ \bibnamefont
  {Campbell}}, \bibinfo {author} {\bibfnamefont {I.~S.}\ \bibnamefont {Heng}},
  \bibinfo {author} {\bibfnamefont {S.}~\bibnamefont {Galliou}}, \bibinfo
  {author} {\bibfnamefont {E.~N.}\ \bibnamefont {Ivanov}},\ and\ \bibinfo
  {author} {\bibfnamefont {M.~E.}\ \bibnamefont {Tobar}},\ }\bibfield  {title}
  {\bibinfo {title} {Rare events detected with a bulk acoustic wave high
  frequency gravitational wave antenna},\ }\href
  {https://doi.org/10.1103/PhysRevLett.127.071102} {\bibfield  {journal}
  {\bibinfo  {journal} {Phys. Rev. Lett.}\ }\textbf {\bibinfo {volume} {127}},\
  \bibinfo {pages} {071102} (\bibinfo {year} {2021})}\BibitemShut {NoStop}%
\bibitem [{\citenamefont {Campbell}\ \emph {et~al.}(2021)\citenamefont
  {Campbell}, \citenamefont {McAllister}, \citenamefont {Goryachev},
  \citenamefont {Ivanov},\ and\ \citenamefont {Tobar}}]{Campbell2021}%
  \BibitemOpen
  \bibfield  {author} {\bibinfo {author} {\bibfnamefont {W.~M.}\ \bibnamefont
  {Campbell}}, \bibinfo {author} {\bibfnamefont {B.~T.}\ \bibnamefont
  {McAllister}}, \bibinfo {author} {\bibfnamefont {M.}~\bibnamefont
  {Goryachev}}, \bibinfo {author} {\bibfnamefont {E.~N.}\ \bibnamefont
  {Ivanov}},\ and\ \bibinfo {author} {\bibfnamefont {M.~E.}\ \bibnamefont
  {Tobar}},\ }\bibfield  {title} {\bibinfo {title} {Searching for scalar dark
  matter via coupling to fundamental constants with photonic, atomic, and
  mechanical oscillators},\ }\href
  {https://doi.org/10.1103/physrevlett.126.071301} {\bibfield  {journal}
  {\bibinfo  {journal} {Phys, Rev. Lett.}\ }\textbf {\bibinfo {volume} {126}},\
  \bibinfo {pages} {071301} (\bibinfo {year} {2021})}\BibitemShut {NoStop}%
\bibitem [{\citenamefont {Campbell}\ \emph
  {et~al.}(2023{\natexlab{a}})\citenamefont {Campbell}, \citenamefont
  {Goryachev},\ and\ \citenamefont {Tobar}}]{Campbell2023mage}%
  \BibitemOpen
  \bibfield  {author} {\bibinfo {author} {\bibfnamefont {W.~M.}\ \bibnamefont
  {Campbell}}, \bibinfo {author} {\bibfnamefont {M.}~\bibnamefont
  {Goryachev}},\ and\ \bibinfo {author} {\bibfnamefont {M.~E.}\ \bibnamefont
  {Tobar}},\ }\bibfield  {title} {\bibinfo {title} {The multi-mode acoustic
  gravitational wave experiment: Mage},\ }\href
  {https://doi.org/10.1038/s41598-023-35670-y} {\bibfield  {journal} {\bibinfo
  {journal} {Scientific Reports}\ }\textbf {\bibinfo {volume} {13}},\ \bibinfo
  {pages} {10638} (\bibinfo {year} {2023}{\natexlab{a}})}\BibitemShut {NoStop}%
\bibitem [{\citenamefont {Campbell}\ \emph
  {et~al.}(2023{\natexlab{b}})\citenamefont {Campbell}, \citenamefont {Tobar},
  \citenamefont {Goryachev},\ and\ \citenamefont {Galliou}}]{Campbell2023}%
  \BibitemOpen
  \bibfield  {author} {\bibinfo {author} {\bibfnamefont {W.~M.}\ \bibnamefont
  {Campbell}}, \bibinfo {author} {\bibfnamefont {M.~E.}\ \bibnamefont {Tobar}},
  \bibinfo {author} {\bibfnamefont {M.}~\bibnamefont {Goryachev}},\ and\
  \bibinfo {author} {\bibfnamefont {S.}~\bibnamefont {Galliou}},\ }\bibfield
  {title} {\bibinfo {title} {Improved constraints on minimum length models with
  a macroscopic low loss phonon cavity},\ }\href
  {https://doi.org/10.1103/PhysRevD.108.102006} {\bibfield  {journal} {\bibinfo
   {journal} {Phys. Rev. D}\ }\textbf {\bibinfo {volume} {108}},\ \bibinfo
  {pages} {102006} (\bibinfo {year} {2023}{\natexlab{b}})}\BibitemShut
  {NoStop}%
\bibitem [{\citenamefont {Aggarwal}\ \emph {et~al.}(2021)\citenamefont
  {Aggarwal}, \citenamefont {Aguiar}, \citenamefont {Bauswein}, \citenamefont
  {Cella}, \citenamefont {Clesse}, \citenamefont {Cruise}, \citenamefont
  {Domcke}, \citenamefont {Figueroa}, \citenamefont {Geraci}, \citenamefont
  {Goryachev}, \citenamefont {Grote}, \citenamefont {Hindmarsh}, \citenamefont
  {Muia}, \citenamefont {Mukund}, \citenamefont {Ottaway}, \citenamefont
  {Peloso}, \citenamefont {Quevedo}, \citenamefont {Ricciardone}, \citenamefont
  {Steinlechner}, \citenamefont {Steinlechner}, \citenamefont {Sun},
  \citenamefont {Tobar}, \citenamefont {Torrenti}, \citenamefont {{\"{U}}nal},\
  and\ \citenamefont {White}}]{Aggarwal2021}%
  \BibitemOpen
  \bibfield  {author} {\bibinfo {author} {\bibfnamefont {N.}~\bibnamefont
  {Aggarwal}}, \bibinfo {author} {\bibfnamefont {O.~D.}\ \bibnamefont
  {Aguiar}}, \bibinfo {author} {\bibfnamefont {A.}~\bibnamefont {Bauswein}},
  \bibinfo {author} {\bibfnamefont {G.}~\bibnamefont {Cella}}, \bibinfo
  {author} {\bibfnamefont {S.}~\bibnamefont {Clesse}}, \bibinfo {author}
  {\bibfnamefont {A.~M.}\ \bibnamefont {Cruise}}, \bibinfo {author}
  {\bibfnamefont {V.}~\bibnamefont {Domcke}}, \bibinfo {author} {\bibfnamefont
  {D.~G.}\ \bibnamefont {Figueroa}}, \bibinfo {author} {\bibfnamefont
  {A.}~\bibnamefont {Geraci}}, \bibinfo {author} {\bibfnamefont
  {M.}~\bibnamefont {Goryachev}}, \bibinfo {author} {\bibfnamefont
  {H.}~\bibnamefont {Grote}}, \bibinfo {author} {\bibfnamefont
  {M.}~\bibnamefont {Hindmarsh}}, \bibinfo {author} {\bibfnamefont
  {F.}~\bibnamefont {Muia}}, \bibinfo {author} {\bibfnamefont {N.}~\bibnamefont
  {Mukund}}, \bibinfo {author} {\bibfnamefont {D.}~\bibnamefont {Ottaway}},
  \bibinfo {author} {\bibfnamefont {M.}~\bibnamefont {Peloso}}, \bibinfo
  {author} {\bibfnamefont {F.}~\bibnamefont {Quevedo}}, \bibinfo {author}
  {\bibfnamefont {A.}~\bibnamefont {Ricciardone}}, \bibinfo {author}
  {\bibfnamefont {J.}~\bibnamefont {Steinlechner}}, \bibinfo {author}
  {\bibfnamefont {S.}~\bibnamefont {Steinlechner}}, \bibinfo {author}
  {\bibfnamefont {S.}~\bibnamefont {Sun}}, \bibinfo {author} {\bibfnamefont
  {M.~E.}\ \bibnamefont {Tobar}}, \bibinfo {author} {\bibfnamefont
  {F.}~\bibnamefont {Torrenti}}, \bibinfo {author} {\bibfnamefont
  {C.}~\bibnamefont {{\"{U}}nal}},\ and\ \bibinfo {author} {\bibfnamefont
  {G.}~\bibnamefont {White}},\ }\href
  {https://doi.org/10.1007/s41114-021-00032-5} {\emph {\bibinfo {title} {Living
  Reviews in Relativity}}},\ Vol.~\bibinfo {volume} {24}\ (\bibinfo {year}
  {2021})\ \Eprint {https://arxiv.org/abs/2011.12414} {arXiv:2011.12414}
  \BibitemShut {NoStop}%
\bibitem [{\citenamefont {Goryachev}\ \emph {et~al.}(2018)\citenamefont
  {Goryachev}, \citenamefont {Kuang}, \citenamefont {Ivanov}, \citenamefont
  {Haslinger}, \citenamefont {Muller},\ and\ \citenamefont
  {Tobar}}]{Goryachev2018}%
  \BibitemOpen
  \bibfield  {author} {\bibinfo {author} {\bibfnamefont {M.}~\bibnamefont
  {Goryachev}}, \bibinfo {author} {\bibfnamefont {Z.}~\bibnamefont {Kuang}},
  \bibinfo {author} {\bibfnamefont {E.~N.}\ \bibnamefont {Ivanov}}, \bibinfo
  {author} {\bibfnamefont {P.}~\bibnamefont {Haslinger}}, \bibinfo {author}
  {\bibfnamefont {H.}~\bibnamefont {Muller}},\ and\ \bibinfo {author}
  {\bibfnamefont {M.~E.}\ \bibnamefont {Tobar}},\ }\bibfield  {title} {\bibinfo
  {title} {Next generation of phonon tests of lorentz invariance using quartz
  {BAW} resonators},\ }\href {https://doi.org/10.1109/tuffc.2018.2824845}
  {\bibfield  {journal} {\bibinfo  {journal} {{IEEE} Transactions on
  Ultrasonics, Ferroelectrics, and Frequency Control}\ }\textbf {\bibinfo
  {volume} {65}},\ \bibinfo {pages} {991} (\bibinfo {year} {2018})}\BibitemShut
  {NoStop}%
\bibitem [{\citenamefont {Lo}\ \emph {et~al.}(2016)\citenamefont {Lo},
  \citenamefont {Haslinger}, \citenamefont {Mizrachi}, \citenamefont
  {Anderegg}, \citenamefont {M\"uller}, \citenamefont {Hohensee}, \citenamefont
  {Goryachev},\ and\ \citenamefont {Tobar}}]{Lo2016}%
  \BibitemOpen
  \bibfield  {author} {\bibinfo {author} {\bibfnamefont {A.}~\bibnamefont
  {Lo}}, \bibinfo {author} {\bibfnamefont {P.}~\bibnamefont {Haslinger}},
  \bibinfo {author} {\bibfnamefont {E.}~\bibnamefont {Mizrachi}}, \bibinfo
  {author} {\bibfnamefont {L.}~\bibnamefont {Anderegg}}, \bibinfo {author}
  {\bibfnamefont {H.}~\bibnamefont {M\"uller}}, \bibinfo {author}
  {\bibfnamefont {M.}~\bibnamefont {Hohensee}}, \bibinfo {author}
  {\bibfnamefont {M.}~\bibnamefont {Goryachev}},\ and\ \bibinfo {author}
  {\bibfnamefont {M.~E.}\ \bibnamefont {Tobar}},\ }\bibfield  {title} {\bibinfo
  {title} {Acoustic tests of lorentz symmetry using quartz oscillators},\
  }\href {https://doi.org/10.1103/PhysRevX.6.011018} {\bibfield  {journal}
  {\bibinfo  {journal} {Phys. Rev. X}\ }\textbf {\bibinfo {volume} {6}},\
  \bibinfo {pages} {011018} (\bibinfo {year} {2016})}\BibitemShut {NoStop}%
\bibitem [{\citenamefont {Galliou}\ \emph {et~al.}(2013)\citenamefont
  {Galliou}, \citenamefont {Goryachev}, \citenamefont {Bourquin}, \citenamefont
  {Abbe}, \citenamefont {Aubry},\ and\ \citenamefont {Tobar}}]{ScRep}%
  \BibitemOpen
  \bibfield  {author} {\bibinfo {author} {\bibfnamefont {S.}~\bibnamefont
  {Galliou}}, \bibinfo {author} {\bibfnamefont {M.}~\bibnamefont {Goryachev}},
  \bibinfo {author} {\bibfnamefont {R.}~\bibnamefont {Bourquin}}, \bibinfo
  {author} {\bibfnamefont {P.}~\bibnamefont {Abbe}}, \bibinfo {author}
  {\bibfnamefont {J.}~\bibnamefont {Aubry}},\ and\ \bibinfo {author}
  {\bibfnamefont {M.}~\bibnamefont {Tobar}},\ }\bibfield  {title} {\bibinfo
  {title} {Extremely low loss phonon-trapping cryogenic acoustic cavities for
  future physical experiments},\ }\href@noop {} {\bibfield  {journal} {\bibinfo
   {journal} {Nature: Scientific Reports}\ }\textbf {\bibinfo {volume} {3}}
  (\bibinfo {year} {2013})}\BibitemShut {NoStop}%
\bibitem [{\citenamefont {Goryachev}\ \emph
  {et~al.}(2013{\natexlab{a}})\citenamefont {Goryachev}, \citenamefont
  {Galliou}, \citenamefont {Imbaud},\ and\ \citenamefont
  {Abb{\'e}}}]{Goryachev:2013ly}%
  \BibitemOpen
  \bibfield  {author} {\bibinfo {author} {\bibfnamefont {M.}~\bibnamefont
  {Goryachev}}, \bibinfo {author} {\bibfnamefont {S.}~\bibnamefont {Galliou}},
  \bibinfo {author} {\bibfnamefont {J.}~\bibnamefont {Imbaud}},\ and\ \bibinfo
  {author} {\bibfnamefont {P.}~\bibnamefont {Abb{\'e}}},\ }\bibfield  {title}
  {\bibinfo {title} {Advances in development of quartz crystal oscillators at
  liquid helium temperatures},\ }\href
  {https://doi.org/http://dx.doi.org/10.1016/j.cryogenics.2013.06.001}
  {\bibfield  {journal} {\bibinfo  {journal} {Cryogenics}\ }\textbf {\bibinfo
  {volume} {57}},\ \bibinfo {pages} {104} (\bibinfo {year}
  {2013}{\natexlab{a}})}\BibitemShut {NoStop}%
\bibitem [{\citenamefont {Galliou}\ \emph {et~al.}(2011)\citenamefont
  {Galliou}, \citenamefont {Imbaud}, \citenamefont {Goryachev}, \citenamefont
  {Bourquin},\ and\ \citenamefont {Abb{\'e}}}]{apl1}%
  \BibitemOpen
  \bibfield  {author} {\bibinfo {author} {\bibfnamefont {S.}~\bibnamefont
  {Galliou}}, \bibinfo {author} {\bibfnamefont {J.}~\bibnamefont {Imbaud}},
  \bibinfo {author} {\bibfnamefont {M.}~\bibnamefont {Goryachev}}, \bibinfo
  {author} {\bibfnamefont {R.}~\bibnamefont {Bourquin}},\ and\ \bibinfo
  {author} {\bibfnamefont {P.}~\bibnamefont {Abb{\'e}}},\ }\bibfield  {title}
  {\bibinfo {title} {Losses in high quality quartz crystal resonators at
  cryogenic temperatures},\ }\href
  {https://doi.org/http://dx.doi.org/10.1063/1.3559611} {\bibfield  {journal}
  {\bibinfo  {journal} {Applied Physics Letters}\ }\textbf {\bibinfo {volume}
  {98}},\ \bibinfo {eid} {091911} (\bibinfo {year} {2011})}\BibitemShut
  {NoStop}%
\bibitem [{\citenamefont {Goryachev}\ \emph {et~al.}(2012)\citenamefont
  {Goryachev}, \citenamefont {Creedon}, \citenamefont {Ivanov}, \citenamefont
  {Galliou}, \citenamefont {Bourquin},\ and\ \citenamefont {Tobar}}]{apl2}%
  \BibitemOpen
  \bibfield  {author} {\bibinfo {author} {\bibfnamefont {M.}~\bibnamefont
  {Goryachev}}, \bibinfo {author} {\bibfnamefont {D.~L.}\ \bibnamefont
  {Creedon}}, \bibinfo {author} {\bibfnamefont {E.~N.}\ \bibnamefont {Ivanov}},
  \bibinfo {author} {\bibfnamefont {S.}~\bibnamefont {Galliou}}, \bibinfo
  {author} {\bibfnamefont {R.}~\bibnamefont {Bourquin}},\ and\ \bibinfo
  {author} {\bibfnamefont {M.~E.}\ \bibnamefont {Tobar}},\ }\bibfield  {title}
  {\bibinfo {title} {Extremely low-loss acoustic phonons in a quartz bulk
  acoustic wave resonator at millikelvin temperature},\ }\href
  {https://doi.org/http://dx.doi.org/10.1063/1.4729292} {\bibfield  {journal}
  {\bibinfo  {journal} {Applied Physics Letters}\ }\textbf {\bibinfo {volume}
  {100}},\ \bibinfo {eid} {243504} (\bibinfo {year} {2012})}\BibitemShut
  {NoStop}%
\bibitem [{\citenamefont {Guarino}\ \emph {et~al.}(2007)\citenamefont
  {Guarino}, \citenamefont {Poberaj}, \citenamefont {Rezzonico}, \citenamefont
  {Degl'Innocenti},\ and\ \citenamefont {G{\"u}nter}}]{Guarino2007}%
  \BibitemOpen
  \bibfield  {author} {\bibinfo {author} {\bibfnamefont {A.}~\bibnamefont
  {Guarino}}, \bibinfo {author} {\bibfnamefont {G.}~\bibnamefont {Poberaj}},
  \bibinfo {author} {\bibfnamefont {D.}~\bibnamefont {Rezzonico}}, \bibinfo
  {author} {\bibfnamefont {R.}~\bibnamefont {Degl'Innocenti}},\ and\ \bibinfo
  {author} {\bibfnamefont {P.}~\bibnamefont {G{\"u}nter}},\ }\bibfield  {title}
  {\bibinfo {title} {Electro--optically tunable microring resonators in lithium
  niobate},\ }\href {https://doi.org/10.1038/nphoton.2007.93} {\bibfield
  {journal} {\bibinfo  {journal} {Nature Photonics}\ }\textbf {\bibinfo
  {volume} {1}},\ \bibinfo {pages} {407} (\bibinfo {year} {2007})}\BibitemShut
  {NoStop}%
\bibitem [{\citenamefont {Zhang}\ \emph {et~al.}(2017)\citenamefont {Zhang},
  \citenamefont {Wang}, \citenamefont {Cheng}, \citenamefont {Shams-Ansari},\
  and\ \citenamefont {Lon\v{c}ar}}]{Zhang:17}%
  \BibitemOpen
  \bibfield  {author} {\bibinfo {author} {\bibfnamefont {M.}~\bibnamefont
  {Zhang}}, \bibinfo {author} {\bibfnamefont {C.}~\bibnamefont {Wang}},
  \bibinfo {author} {\bibfnamefont {R.}~\bibnamefont {Cheng}}, \bibinfo
  {author} {\bibfnamefont {A.}~\bibnamefont {Shams-Ansari}},\ and\ \bibinfo
  {author} {\bibfnamefont {M.}~\bibnamefont {Lon\v{c}ar}},\ }\bibfield  {title}
  {\bibinfo {title} {Monolithic ultra-high-q lithium niobate microring
  resonator},\ }\href {https://doi.org/10.1364/OPTICA.4.001536} {\bibfield
  {journal} {\bibinfo  {journal} {Optica}\ }\textbf {\bibinfo {volume} {4}},\
  \bibinfo {pages} {1536} (\bibinfo {year} {2017})}\BibitemShut {NoStop}%
\bibitem [{\citenamefont {Zhang}\ \emph {et~al.}(2019)\citenamefont {Zhang},
  \citenamefont {Buscaino}, \citenamefont {Wang}, \citenamefont {Shams-Ansari},
  \citenamefont {Reimer}, \citenamefont {Zhu}, \citenamefont {Kahn},\ and\
  \citenamefont {Lon{\v c}ar}}]{Zhang2019}%
  \BibitemOpen
  \bibfield  {author} {\bibinfo {author} {\bibfnamefont {M.}~\bibnamefont
  {Zhang}}, \bibinfo {author} {\bibfnamefont {B.}~\bibnamefont {Buscaino}},
  \bibinfo {author} {\bibfnamefont {C.}~\bibnamefont {Wang}}, \bibinfo {author}
  {\bibfnamefont {A.}~\bibnamefont {Shams-Ansari}}, \bibinfo {author}
  {\bibfnamefont {C.}~\bibnamefont {Reimer}}, \bibinfo {author} {\bibfnamefont
  {R.}~\bibnamefont {Zhu}}, \bibinfo {author} {\bibfnamefont {J.~M.}\
  \bibnamefont {Kahn}},\ and\ \bibinfo {author} {\bibfnamefont
  {M.}~\bibnamefont {Lon{\v c}ar}},\ }\bibfield  {title} {\bibinfo {title}
  {Broadband electro-optic frequency comb generation in a lithium niobate
  microring resonator},\ }\href {https://doi.org/10.1038/s41586-019-1008-7}
  {\bibfield  {journal} {\bibinfo  {journal} {Nature}\ }\textbf {\bibinfo
  {volume} {568}},\ \bibinfo {pages} {373} (\bibinfo {year}
  {2019})}\BibitemShut {NoStop}%
\bibitem [{\citenamefont {Gong}\ and\ \citenamefont {Piazza}(2013)}]{Gong2013}%
  \BibitemOpen
  \bibfield  {author} {\bibinfo {author} {\bibfnamefont {S.}~\bibnamefont
  {Gong}}\ and\ \bibinfo {author} {\bibfnamefont {G.}~\bibnamefont {Piazza}},\
  }\bibfield  {title} {\bibinfo {title} {Design and analysis of
  lithium--niobate-based high electromechanical coupling rf-mems resonators for
  wideband filtering},\ }\href {https://doi.org/10.1109/TMTT.2012.2228671}
  {\bibfield  {journal} {\bibinfo  {journal} {IEEE Transactions on Microwave
  Theory and Techniques}\ }\textbf {\bibinfo {volume} {61}},\ \bibinfo {pages}
  {403} (\bibinfo {year} {2013})}\BibitemShut {NoStop}%
\bibitem [{\citenamefont {Marinkovi{\'c}}\ \emph {et~al.}(2021)\citenamefont
  {Marinkovi{\'c}}, \citenamefont {Drimmer}, \citenamefont {Hensen},\ and\
  \citenamefont {Gr{\"o}blacher}}]{Marink2021}%
  \BibitemOpen
  \bibfield  {author} {\bibinfo {author} {\bibfnamefont {I.}~\bibnamefont
  {Marinkovi{\'c}}}, \bibinfo {author} {\bibfnamefont {M.}~\bibnamefont
  {Drimmer}}, \bibinfo {author} {\bibfnamefont {B.}~\bibnamefont {Hensen}},\
  and\ \bibinfo {author} {\bibfnamefont {S.}~\bibnamefont {Gr{\"o}blacher}},\
  }\bibfield  {title} {\bibinfo {title} {Hybrid integration of silicon photonic
  devices on lithium niobate for optomechanical wavelength conversion},\ }\href
  {https://doi.org/10.1021/acs.nanolett.0c03980} {\bibfield  {journal}
  {\bibinfo  {journal} {Nano Letters}\ }\textbf {\bibinfo {volume} {21}},\
  \bibinfo {pages} {529} (\bibinfo {year} {2021})}\BibitemShut {NoStop}%
\bibitem [{\citenamefont {Shen}\ \emph {et~al.}(2020)\citenamefont {Shen},
  \citenamefont {Xie}, \citenamefont {Zou}, \citenamefont {Xu}, \citenamefont
  {Fu},\ and\ \citenamefont {Tang}}]{Shen2020}%
  \BibitemOpen
  \bibfield  {author} {\bibinfo {author} {\bibfnamefont {M.}~\bibnamefont
  {Shen}}, \bibinfo {author} {\bibfnamefont {J.}~\bibnamefont {Xie}}, \bibinfo
  {author} {\bibfnamefont {C.-L.}\ \bibnamefont {Zou}}, \bibinfo {author}
  {\bibfnamefont {Y.}~\bibnamefont {Xu}}, \bibinfo {author} {\bibfnamefont
  {W.}~\bibnamefont {Fu}},\ and\ \bibinfo {author} {\bibfnamefont {H.~X.}\
  \bibnamefont {Tang}},\ }\bibfield  {title} {\bibinfo {title} {{High frequency
  lithium niobate film-thickness-mode optomechanical resonator}},\ }\href
  {https://doi.org/10.1063/5.0020019} {\bibfield  {journal} {\bibinfo
  {journal} {Applied Physics Letters}\ }\textbf {\bibinfo {volume} {117}},\
  \bibinfo {pages} {131104} (\bibinfo {year} {2020})},\ \Eprint
  {https://arxiv.org/abs/https://pubs.aip.org/aip/apl/article-pdf/doi/10.1063/5.0020019/14538519/131104\_1\_online.pdf}
  {https://pubs.aip.org/aip/apl/article-pdf/doi/10.1063/5.0020019/14538519/131104\_1\_online.pdf}
  \BibitemShut {NoStop}%
\bibitem [{\citenamefont {Zorzetti}\ \emph {et~al.}(2023)\citenamefont
  {Zorzetti}, \citenamefont {Wang}, \citenamefont {Gonin}, \citenamefont
  {Kazakov}, \citenamefont {Khabiboulline}, \citenamefont {Romanenko},
  \citenamefont {Yakovlev},\ and\ \citenamefont {Grassellino}}]{Zorzetti2023}%
  \BibitemOpen
  \bibfield  {author} {\bibinfo {author} {\bibfnamefont {S.}~\bibnamefont
  {Zorzetti}}, \bibinfo {author} {\bibfnamefont {C.}~\bibnamefont {Wang}},
  \bibinfo {author} {\bibfnamefont {I.}~\bibnamefont {Gonin}}, \bibinfo
  {author} {\bibfnamefont {S.}~\bibnamefont {Kazakov}}, \bibinfo {author}
  {\bibfnamefont {T.}~\bibnamefont {Khabiboulline}}, \bibinfo {author}
  {\bibfnamefont {A.}~\bibnamefont {Romanenko}}, \bibinfo {author}
  {\bibfnamefont {V.~P.}\ \bibnamefont {Yakovlev}},\ and\ \bibinfo {author}
  {\bibfnamefont {A.}~\bibnamefont {Grassellino}},\ }\bibfield  {title}
  {\bibinfo {title} {Millikelvin measurements of permittivity and loss tangent
  of lithium niobate},\ }\href {https://doi.org/10.1103/PhysRevB.107.L220302}
  {\bibfield  {journal} {\bibinfo  {journal} {Phys. Rev. B}\ }\textbf {\bibinfo
  {volume} {107}},\ \bibinfo {pages} {L220302} (\bibinfo {year}
  {2023})}\BibitemShut {NoStop}%
\bibitem [{\citenamefont {Wang}\ \emph {et~al.}(2022)\citenamefont {Wang},
  \citenamefont {Gonin}, \citenamefont {Grassellino}, \citenamefont {Kazakov},
  \citenamefont {Romanenko}, \citenamefont {Yakovlev},\ and\ \citenamefont
  {Zorzetti}}]{Wang2022}%
  \BibitemOpen
  \bibfield  {author} {\bibinfo {author} {\bibfnamefont {C.}~\bibnamefont
  {Wang}}, \bibinfo {author} {\bibfnamefont {I.}~\bibnamefont {Gonin}},
  \bibinfo {author} {\bibfnamefont {A.}~\bibnamefont {Grassellino}}, \bibinfo
  {author} {\bibfnamefont {S.}~\bibnamefont {Kazakov}}, \bibinfo {author}
  {\bibfnamefont {A.}~\bibnamefont {Romanenko}}, \bibinfo {author}
  {\bibfnamefont {V.~P.}\ \bibnamefont {Yakovlev}},\ and\ \bibinfo {author}
  {\bibfnamefont {S.}~\bibnamefont {Zorzetti}},\ }\bibfield  {title} {\bibinfo
  {title} {High-efficiency microwave-optical quantum transduction based on a
  cavity electro-optic superconducting system with long coherence time},\
  }\href {https://doi.org/10.1038/s41534-022-00664-7} {\bibfield  {journal}
  {\bibinfo  {journal} {npj Quantum Information}\ }\textbf {\bibinfo {volume}
  {8}},\ \bibinfo {pages} {149} (\bibinfo {year} {2022})}\BibitemShut {NoStop}%
\bibitem [{\citenamefont {Parashar}\ \emph {et~al.}(2024)\citenamefont
  {Parashar}, \citenamefont {Campbell}, \citenamefont {Bourhill}, \citenamefont
  {Ivanov}, \citenamefont {Goryachev},\ and\ \citenamefont
  {Tobar}}]{10.1063/5.0233800}%
  \BibitemOpen
  \bibfield  {author} {\bibinfo {author} {\bibfnamefont {S.}~\bibnamefont
  {Parashar}}, \bibinfo {author} {\bibfnamefont {W.~M.}\ \bibnamefont
  {Campbell}}, \bibinfo {author} {\bibfnamefont {J.}~\bibnamefont {Bourhill}},
  \bibinfo {author} {\bibfnamefont {E.}~\bibnamefont {Ivanov}}, \bibinfo
  {author} {\bibfnamefont {M.}~\bibnamefont {Goryachev}},\ and\ \bibinfo
  {author} {\bibfnamefont {M.~E.}\ \bibnamefont {Tobar}},\ }\bibfield  {title}
  {\bibinfo {title} {Upconversion of phonon modes into microwave photons in a
  lithium niobate bulk acoustic wave resonator coupled to a microwave cavity},\
  }\href {https://doi.org/10.1063/5.0233800} {\bibfield  {journal} {\bibinfo
  {journal} {APL Photonics}\ }\textbf {\bibinfo {volume} {9}},\ \bibinfo
  {pages} {111304} (\bibinfo {year} {2024})},\ \Eprint
  {https://arxiv.org/abs/https://pubs.aip.org/aip/app/article-pdf/doi/10.1063/5.0233800/20250374/111304\_1\_5.0233800.pdf}
  {https://pubs.aip.org/aip/app/article-pdf/doi/10.1063/5.0233800/20250374/111304\_1\_5.0233800.pdf}
  \BibitemShut {NoStop}%
\bibitem [{\citenamefont {Wollack}\ \emph {et~al.}(2021)\citenamefont
  {Wollack}, \citenamefont {Cleland}, \citenamefont {Arrangoiz-Arriola},
  \citenamefont {McKenna}, \citenamefont {Gruenke}, \citenamefont {Patel},
  \citenamefont {Jiang}, \citenamefont {Sarabalis},\ and\ \citenamefont
  {Safavi-Naeini}}]{Wollack2021}%
  \BibitemOpen
  \bibfield  {author} {\bibinfo {author} {\bibfnamefont {E.~A.}\ \bibnamefont
  {Wollack}}, \bibinfo {author} {\bibfnamefont {A.~Y.}\ \bibnamefont
  {Cleland}}, \bibinfo {author} {\bibfnamefont {P.}~\bibnamefont
  {Arrangoiz-Arriola}}, \bibinfo {author} {\bibfnamefont {T.~P.}\ \bibnamefont
  {McKenna}}, \bibinfo {author} {\bibfnamefont {R.~G.}\ \bibnamefont
  {Gruenke}}, \bibinfo {author} {\bibfnamefont {R.~N.}\ \bibnamefont {Patel}},
  \bibinfo {author} {\bibfnamefont {W.}~\bibnamefont {Jiang}}, \bibinfo
  {author} {\bibfnamefont {C.~J.}\ \bibnamefont {Sarabalis}},\ and\ \bibinfo
  {author} {\bibfnamefont {A.~H.}\ \bibnamefont {Safavi-Naeini}},\ }\bibfield
  {title} {\bibinfo {title} {{Loss channels affecting lithium niobate phononic
  crystal resonators at cryogenic temperature}},\ }\href
  {https://doi.org/10.1063/5.0034909} {\bibfield  {journal} {\bibinfo
  {journal} {Applied Physics Letters}\ }\textbf {\bibinfo {volume} {118}},\
  \bibinfo {pages} {123501} (\bibinfo {year} {2021})},\ \Eprint
  {https://arxiv.org/abs/https://pubs.aip.org/aip/apl/article-pdf/doi/10.1063/5.0034909/14546738/123501\_1\_online.pdf}
  {https://pubs.aip.org/aip/apl/article-pdf/doi/10.1063/5.0034909/14546738/123501\_1\_online.pdf}
  \BibitemShut {NoStop}%
\bibitem [{\citenamefont {Kitzman}\ \emph
  {et~al.}(2023{\natexlab{a}})\citenamefont {Kitzman}, \citenamefont {Lane},
  \citenamefont {Undershute}, \citenamefont {Drimmer}, \citenamefont
  {Schleusner}, \citenamefont {Beysengulov}, \citenamefont {Mikolas},\ and\
  \citenamefont {Pollanen}}]{Kitzman2023}%
  \BibitemOpen
  \bibfield  {author} {\bibinfo {author} {\bibfnamefont {J.~M.}\ \bibnamefont
  {Kitzman}}, \bibinfo {author} {\bibfnamefont {J.~R.}\ \bibnamefont {Lane}},
  \bibinfo {author} {\bibfnamefont {C.}~\bibnamefont {Undershute}}, \bibinfo
  {author} {\bibfnamefont {M.}~\bibnamefont {Drimmer}}, \bibinfo {author}
  {\bibfnamefont {A.~J.}\ \bibnamefont {Schleusner}}, \bibinfo {author}
  {\bibfnamefont {N.~R.}\ \bibnamefont {Beysengulov}}, \bibinfo {author}
  {\bibfnamefont {C.~A.}\ \bibnamefont {Mikolas}},\ and\ \bibinfo {author}
  {\bibfnamefont {J.}~\bibnamefont {Pollanen}},\ }\bibfield  {title} {\bibinfo
  {title} {{Free-space coupling and characterization of transverse bulk phonon
  modes in lithium niobate in a quantum acoustic device}},\ }\href
  {https://doi.org/10.1063/5.0170221} {\bibfield  {journal} {\bibinfo
  {journal} {Applied Physics Letters}\ }\textbf {\bibinfo {volume} {123}},\
  \bibinfo {pages} {224001} (\bibinfo {year} {2023}{\natexlab{a}})},\ \Eprint
  {https://arxiv.org/abs/https://pubs.aip.org/aip/apl/article-pdf/doi/10.1063/5.0170221/18226262/224001\_1\_5.0170221.pdf}
  {https://pubs.aip.org/aip/apl/article-pdf/doi/10.1063/5.0170221/18226262/224001\_1\_5.0170221.pdf}
  \BibitemShut {NoStop}%
\bibitem [{\citenamefont {Kitzman}\ \emph
  {et~al.}(2023{\natexlab{b}})\citenamefont {Kitzman}, \citenamefont {Lane},
  \citenamefont {Undershute}, \citenamefont {Harrington}, \citenamefont
  {Beysengulov}, \citenamefont {Mikolas}, \citenamefont {Murch},\ and\
  \citenamefont {Pollanen}}]{Kitzman2023b}%
  \BibitemOpen
  \bibfield  {author} {\bibinfo {author} {\bibfnamefont {J.~M.}\ \bibnamefont
  {Kitzman}}, \bibinfo {author} {\bibfnamefont {J.~R.}\ \bibnamefont {Lane}},
  \bibinfo {author} {\bibfnamefont {C.}~\bibnamefont {Undershute}}, \bibinfo
  {author} {\bibfnamefont {P.~M.}\ \bibnamefont {Harrington}}, \bibinfo
  {author} {\bibfnamefont {N.~R.}\ \bibnamefont {Beysengulov}}, \bibinfo
  {author} {\bibfnamefont {C.~A.}\ \bibnamefont {Mikolas}}, \bibinfo {author}
  {\bibfnamefont {K.~W.}\ \bibnamefont {Murch}},\ and\ \bibinfo {author}
  {\bibfnamefont {J.}~\bibnamefont {Pollanen}},\ }\bibfield  {title} {\bibinfo
  {title} {Phononic bath engineering of a superconducting qubit},\ }\href
  {https://doi.org/10.1038/s41467-023-39682-0} {\bibfield  {journal} {\bibinfo
  {journal} {Nature Communications}\ }\textbf {\bibinfo {volume} {14}},\
  \bibinfo {pages} {3910} (\bibinfo {year} {2023}{\natexlab{b}})}\BibitemShut
  {NoStop}%
\bibitem [{\citenamefont {Goryachev}\ and\ \citenamefont
  {Tobar}(2014{\natexlab{a}})}]{Goryachev:2014aa}%
  \BibitemOpen
  \bibfield  {author} {\bibinfo {author} {\bibfnamefont {M.}~\bibnamefont
  {Goryachev}}\ and\ \bibinfo {author} {\bibfnamefont {M.~E.}\ \bibnamefont
  {Tobar}},\ }\bibfield  {title} {\bibinfo {title} {Effects of geometry on
  quantum fluctuations of phonon-trapping acoustic cavities},\ }\href
  {http://stacks.iop.org/1367-2630/16/i=8/a=083007} {\bibfield  {journal}
  {\bibinfo  {journal} {New Journal of Physics}\ }\textbf {\bibinfo {volume}
  {16}},\ \bibinfo {pages} {083007} (\bibinfo {year}
  {2014}{\natexlab{a}})}\BibitemShut {NoStop}%
\bibitem [{\citenamefont {Goryachev}\ \emph
  {et~al.}(2013{\natexlab{b}})\citenamefont {Goryachev}, \citenamefont
  {Creedon}, \citenamefont {Galliou},\ and\ \citenamefont {Tobar}}]{quartzPRL}%
  \BibitemOpen
  \bibfield  {author} {\bibinfo {author} {\bibfnamefont {M.}~\bibnamefont
  {Goryachev}}, \bibinfo {author} {\bibfnamefont {D.}~\bibnamefont {Creedon}},
  \bibinfo {author} {\bibfnamefont {S.}~\bibnamefont {Galliou}},\ and\ \bibinfo
  {author} {\bibfnamefont {M.}~\bibnamefont {Tobar}},\ }\bibfield  {title}
  {\bibinfo {title} {Observation of rayleigh phonon scattering through
  excitation of extremely high overtones in low-loss cryogenic acoustic
  cavities for hybrid quantum systems},\ }\href@noop {} {\bibfield  {journal}
  {\bibinfo  {journal} {Physical Review Letters}\ }\textbf {\bibinfo {volume}
  {111}},\ \bibinfo {pages} {085502} (\bibinfo {year}
  {2013}{\natexlab{b}})}\BibitemShut {NoStop}%
\bibitem [{\citenamefont {Akheiser}(1939)}]{Akheiser}%
  \BibitemOpen
  \bibfield  {author} {\bibinfo {author} {\bibfnamefont {A.}~\bibnamefont
  {Akheiser}},\ }\bibfield  {title} {\bibinfo {title} {On the absorption of
  sound in solids},\ }\href@noop {} {\bibfield  {journal} {\bibinfo  {journal}
  {Journal of Physics-USSR}\ }\textbf {\bibinfo {volume} {1}} (\bibinfo {year}
  {1939})}\BibitemShut {NoStop}%
\bibitem [{\citenamefont {Landau}\ and\ \citenamefont
  {Rumer}(1937)}]{landaurumer1}%
  \BibitemOpen
  \bibfield  {author} {\bibinfo {author} {\bibfnamefont {L.}~\bibnamefont
  {Landau}}\ and\ \bibinfo {author} {\bibfnamefont {G.}~\bibnamefont {Rumer}},\
  }\bibfield  {title} {\bibinfo {title} {Uber schall absorption in festen
  {K}\"{o}rpen},\ }\href@noop {} {\bibfield  {journal} {\bibinfo  {journal}
  {Physikalische Zeitschrift der Sowjetunion}\ }\textbf {\bibinfo {volume}
  {11}},\ \bibinfo {pages} {18} (\bibinfo {year} {1937})}\BibitemShut {NoStop}%
\bibitem [{\citenamefont {Deresiewicz}(1957)}]{Deresiewicz1957}%
  \BibitemOpen
  \bibfield  {author} {\bibinfo {author} {\bibfnamefont {H.}~\bibnamefont
  {Deresiewicz}},\ }\bibfield  {title} {\bibinfo {title} {Plane waves in a
  thermoelastic solid},\ }\href {https://doi.org/10.1121/1.1908832} {\bibfield
  {journal} {\bibinfo  {journal} {The Journal of the Acoustical Society of
  America}\ }\textbf {\bibinfo {volume} {29}},\ \bibinfo {pages} {204}
  (\bibinfo {year} {1957})},\ \Eprint
  {https://arxiv.org/abs/https://pubs.aip.org/asa/jasa/article-pdf/29/2/204/18736599/204\_1\_online.pdf}
  {https://pubs.aip.org/asa/jasa/article-pdf/29/2/204/18736599/204\_1\_online.pdf}
  \BibitemShut {NoStop}%
\bibitem [{\citenamefont {Goryachev}\ and\ \citenamefont
  {Tobar}(2014{\natexlab{b}})}]{Goryachev:2014ab}%
  \BibitemOpen
  \bibfield  {author} {\bibinfo {author} {\bibfnamefont {M.}~\bibnamefont
  {Goryachev}}\ and\ \bibinfo {author} {\bibfnamefont {M.~E.}\ \bibnamefont
  {Tobar}},\ }\bibfield  {title} {\bibinfo {title} {Gravitational wave
  detection with high frequency phonon trapping acoustic cavities},\ }\href
  {https://doi.org/10.1103/PhysRevD.90.102005} {\bibfield  {journal} {\bibinfo
  {journal} {Physical Review D}\ }\textbf {\bibinfo {volume} {90}},\ \bibinfo
  {pages} {102005} (\bibinfo {year} {2014}{\natexlab{b}})}\BibitemShut
  {NoStop}%
\bibitem [{\citenamefont {w.~Anderson}\ \emph {et~al.}(1972)\citenamefont
  {w.~Anderson}, \citenamefont {Halperin},\ and\ \citenamefont
  {c.~M.~Varma}}]{Anderson1972}%
  \BibitemOpen
  \bibfield  {author} {\bibinfo {author} {\bibfnamefont {P.}~\bibnamefont
  {w.~Anderson}}, \bibinfo {author} {\bibfnamefont {B.~I.}\ \bibnamefont
  {Halperin}},\ and\ \bibinfo {author} {\bibnamefont {c.~M.~Varma}},\
  }\bibfield  {title} {\bibinfo {title} {Anomalous low-temperature thermal
  properties of glasses and spin glasses},\ }\href
  {https://doi.org/10.1080/14786437208229210} {\bibfield  {journal} {\bibinfo
  {journal} {The Philosophical Magazine: A Journal of Theoretical Experimental
  and Applied Physics}\ }\textbf {\bibinfo {volume} {25}},\ \bibinfo {pages}
  {1} (\bibinfo {year} {1972})}\BibitemShut {NoStop}%
\bibitem [{\citenamefont {Phillips}(1972)}]{Phillips1972}%
  \BibitemOpen
  \bibfield  {author} {\bibinfo {author} {\bibfnamefont {W.~A.}\ \bibnamefont
  {Phillips}},\ }\bibfield  {title} {\bibinfo {title} {Tunneling states in
  amorphous solids},\ }\href {https://doi.org/10.1007/bf00660072} {\bibfield
  {journal} {\bibinfo  {journal} {Journal of Low Temperature Physics}\ }\textbf
  {\bibinfo {volume} {7}},\ \bibinfo {pages} {351} (\bibinfo {year}
  {1972})}\BibitemShut {NoStop}%
\bibitem [{\citenamefont {Seoanez}\ \emph {et~al.}(2007)\citenamefont
  {Seoanez}, \citenamefont {Guinea},\ and\ \citenamefont {Neto}}]{Seoanez2007}%
  \BibitemOpen
  \bibfield  {author} {\bibinfo {author} {\bibfnamefont {C.}~\bibnamefont
  {Seoanez}}, \bibinfo {author} {\bibfnamefont {F.}~\bibnamefont {Guinea}},\
  and\ \bibinfo {author} {\bibfnamefont {A.~H.~C.}\ \bibnamefont {Neto}},\
  }\bibfield  {title} {\bibinfo {title} {Dissipation due to two-level systems
  in nano-mechanical devices},\ }\href
  {https://doi.org/10.1209/0295-5075/78/60002} {\bibfield  {journal} {\bibinfo
  {journal} {EPL (Europhysics Letters)}\ }\textbf {\bibinfo {volume} {78}},\
  \bibinfo {pages} {60002} (\bibinfo {year} {2007})}\BibitemShut {NoStop}%
\bibitem [{\citenamefont {Mason}\ and\ \citenamefont
  {McSkimin}(1947)}]{Mason1947}%
  \BibitemOpen
  \bibfield  {author} {\bibinfo {author} {\bibfnamefont {W.~P.}\ \bibnamefont
  {Mason}}\ and\ \bibinfo {author} {\bibfnamefont {H.~J.}\ \bibnamefont
  {McSkimin}},\ }\bibfield  {title} {\bibinfo {title} {Attenuation and
  scattering of high frequency sound waves in metals and glasses},\ }\href
  {https://doi.org/10.1121/1.1916504} {\bibfield  {journal} {\bibinfo
  {journal} {The Journal of the Acoustical Society of America}\ }\textbf
  {\bibinfo {volume} {19}},\ \bibinfo {pages} {464} (\bibinfo {year} {1947})},\
  \Eprint
  {https://arxiv.org/abs/https://pubs.aip.org/asa/jasa/article-pdf/19/3/464/18726994/464\_1\_online.pdf}
  {https://pubs.aip.org/asa/jasa/article-pdf/19/3/464/18726994/464\_1\_online.pdf}
  \BibitemShut {NoStop}%
\bibitem [{\citenamefont {Smith}\ and\ \citenamefont
  {Welsh}(1971)}]{Smith1971}%
  \BibitemOpen
  \bibfield  {author} {\bibinfo {author} {\bibfnamefont {R.~T.}\ \bibnamefont
  {Smith}}\ and\ \bibinfo {author} {\bibfnamefont {F.~S.}\ \bibnamefont
  {Welsh}},\ }\bibfield  {title} {\bibinfo {title} {Temperature dependence of
  the elastic, piezoelectric, and dielectric constants of lithium tantalate and
  lithium niobate},\ }\href {https://doi.org/10.1063/1.1660528} {\bibfield
  {journal} {\bibinfo  {journal} {Journal of Applied Physics}\ }\textbf
  {\bibinfo {volume} {42}},\ \bibinfo {pages} {2219} (\bibinfo {year}
  {1971})},\ \Eprint
  {https://arxiv.org/abs/https://pubs.aip.org/aip/jap/article-pdf/42/6/2219/18357830/2219\_1\_online.pdf}
  {https://pubs.aip.org/aip/jap/article-pdf/42/6/2219/18357830/2219\_1\_online.pdf}
  \BibitemShut {NoStop}%
\bibitem [{\citenamefont {Barron}\ \emph {et~al.}(1982)\citenamefont {Barron},
  \citenamefont {Collins}, \citenamefont {Smith},\ and\ \citenamefont
  {White}}]{Barron1982}%
  \BibitemOpen
  \bibfield  {author} {\bibinfo {author} {\bibfnamefont {T.~H.~K.}\
  \bibnamefont {Barron}}, \bibinfo {author} {\bibfnamefont {J.~F.}\
  \bibnamefont {Collins}}, \bibinfo {author} {\bibfnamefont {T.~W.}\
  \bibnamefont {Smith}},\ and\ \bibinfo {author} {\bibfnamefont {G.~K.}\
  \bibnamefont {White}},\ }\bibfield  {title} {\bibinfo {title} {Thermal
  expansion, gruneisen functions and static lattice properties of quartz},\
  }\href {https://doi.org/10.1088/0022-3719/15/20/016} {\bibfield  {journal}
  {\bibinfo  {journal} {Journal of Physics C: Solid State Physics}\ }\textbf
  {\bibinfo {volume} {15}},\ \bibinfo {pages} {4311} (\bibinfo {year}
  {1982})}\BibitemShut {NoStop}%
\bibitem [{\citenamefont {P\'erez-Enciso}\ and\ \citenamefont
  {Vieira}(1998)}]{Enciso1998}%
  \BibitemOpen
  \bibfield  {author} {\bibinfo {author} {\bibfnamefont {E.}~\bibnamefont
  {P\'erez-Enciso}}\ and\ \bibinfo {author} {\bibfnamefont {S.}~\bibnamefont
  {Vieira}},\ }\bibfield  {title} {\bibinfo {title} {Thermal properties of
  intrinsically disordered ${\mathrm{linbo}}_{3}$ crystals at low
  temperatures},\ }\href {https://doi.org/10.1103/PhysRevB.57.13359} {\bibfield
   {journal} {\bibinfo  {journal} {Phys. Rev. B}\ }\textbf {\bibinfo {volume}
  {57}},\ \bibinfo {pages} {13359} (\bibinfo {year} {1998})}\BibitemShut
  {NoStop}%
\bibitem [{\citenamefont {Gruenke-Freudenstein}\ \emph
  {et~al.}(2025)\citenamefont {Gruenke-Freudenstein}, \citenamefont {Szakiel},
  \citenamefont {Multani}, \citenamefont {Makihara}, \citenamefont {Hayden},
  \citenamefont {Khalatpour}, \citenamefont {Wollack}, \citenamefont
  {Akoto-Yeboah}, \citenamefont {Salmani-Rezaie},\ and\ \citenamefont
  {Safavi-Naeini}}]{gruenkefreudenstein2025}%
  \BibitemOpen
  \bibfield  {author} {\bibinfo {author} {\bibfnamefont {R.~G.}\ \bibnamefont
  {Gruenke-Freudenstein}}, \bibinfo {author} {\bibfnamefont {E.}~\bibnamefont
  {Szakiel}}, \bibinfo {author} {\bibfnamefont {G.~P.}\ \bibnamefont
  {Multani}}, \bibinfo {author} {\bibfnamefont {T.}~\bibnamefont {Makihara}},
  \bibinfo {author} {\bibfnamefont {A.~G.}\ \bibnamefont {Hayden}}, \bibinfo
  {author} {\bibfnamefont {A.}~\bibnamefont {Khalatpour}}, \bibinfo {author}
  {\bibfnamefont {E.~A.}\ \bibnamefont {Wollack}}, \bibinfo {author}
  {\bibfnamefont {A.}~\bibnamefont {Akoto-Yeboah}}, \bibinfo {author}
  {\bibfnamefont {S.}~\bibnamefont {Salmani-Rezaie}},\ and\ \bibinfo {author}
  {\bibfnamefont {A.~H.}\ \bibnamefont {Safavi-Naeini}},\ }\href
  {https://arxiv.org/abs/2501.08291} {\bibinfo {title} {Surface and bulk
  two-level system losses in lithium niobate acoustic resonators}} (\bibinfo
  {year} {2025}),\ \Eprint {https://arxiv.org/abs/2501.08291} {arXiv:2501.08291
  [cond-mat.mes-hall]} \BibitemShut {NoStop}%
\bibitem [{\citenamefont {Imbaud}(2008)}]{imbaud:tel-00360494}%
  \BibitemOpen
  \bibfield  {author} {\bibinfo {author} {\bibfnamefont {J.}~\bibnamefont
  {Imbaud}},\ }\emph {\bibinfo {title} {{Evaluation des potentialit{\'e}s des
  mat{\'e}riaux du type langasite pour la r{\'e}alisation d'oscillateurs
  ultra-stables.<br />Etude et r{\'e}alisation pr{\'e}liminaires d'un
  oscillateur cryog{\'e}nique.}}},\ \href
  {https://theses.hal.science/tel-00360494} {\bibinfo {type} {Theses}},\
  \bibinfo  {school} {{Universit{\'e} de Franche-Comt{\'e}}} (\bibinfo {year}
  {2008})\BibitemShut {NoStop}%
\bibitem [{\citenamefont {Galliou}\ \emph {et~al.}(2014)\citenamefont
  {Galliou}, \citenamefont {Abbe}, \citenamefont {Bourquin}, \citenamefont
  {Goryachev}, \citenamefont {Tobar},\ and\ \citenamefont
  {Ivanov}}]{Galliou2014}%
  \BibitemOpen
  \bibfield  {author} {\bibinfo {author} {\bibfnamefont {S.}~\bibnamefont
  {Galliou}}, \bibinfo {author} {\bibfnamefont {P.}~\bibnamefont {Abbe}},
  \bibinfo {author} {\bibfnamefont {R.}~\bibnamefont {Bourquin}}, \bibinfo
  {author} {\bibfnamefont {M.}~\bibnamefont {Goryachev}}, \bibinfo {author}
  {\bibfnamefont {M.~E.}\ \bibnamefont {Tobar}},\ and\ \bibinfo {author}
  {\bibfnamefont {E.~N.}\ \bibnamefont {Ivanov}},\ }\bibfield  {title}
  {\bibinfo {title} {Properties related to q-factors and noise of quartz
  resonator-based systems at 4k}\ }(\bibinfo  {publisher} {{IEEE}},\ \bibinfo
  {year} {2014})\BibitemShut {NoStop}%
\bibitem [{\citenamefont {Jiang}\ \emph {et~al.}(2004)\citenamefont {Jiang},
  \citenamefont {Yu}, \citenamefont {Liu},\ and\ \citenamefont {Huang}}]{TLS1}%
  \BibitemOpen
  \bibfield  {author} {\bibinfo {author} {\bibfnamefont {H.}~\bibnamefont
  {Jiang}}, \bibinfo {author} {\bibfnamefont {M.-F.}\ \bibnamefont {Yu}},
  \bibinfo {author} {\bibfnamefont {B.}~\bibnamefont {Liu}},\ and\ \bibinfo
  {author} {\bibfnamefont {Y.}~\bibnamefont {Huang}},\ }\bibfield  {title}
  {\bibinfo {title} {Intrinsic energy loss mechanisms in a cantilevered carbon
  nanotube beam oscillator},\ }\href
  {https://doi.org/10.1103/PhysRevLett.93.185501} {\bibfield  {journal}
  {\bibinfo  {journal} {Phys. Rev. Lett.}\ }\textbf {\bibinfo {volume} {93}},\
  \bibinfo {pages} {185501} (\bibinfo {year} {2004})}\BibitemShut {NoStop}%
\bibitem [{\citenamefont {H{\"u}ttel}\ \emph {et~al.}(2009)\citenamefont
  {H{\"u}ttel}, \citenamefont {Steele}, \citenamefont {Witkamp}, \citenamefont
  {Poot}, \citenamefont {Kouwenhoven},\ and\ \citenamefont {van~der
  Zant}}]{TLS2}%
  \BibitemOpen
  \bibfield  {author} {\bibinfo {author} {\bibfnamefont {A.~K.}\ \bibnamefont
  {H{\"u}ttel}}, \bibinfo {author} {\bibfnamefont {G.~A.}\ \bibnamefont
  {Steele}}, \bibinfo {author} {\bibfnamefont {B.}~\bibnamefont {Witkamp}},
  \bibinfo {author} {\bibfnamefont {M.}~\bibnamefont {Poot}}, \bibinfo {author}
  {\bibfnamefont {L.~P.}\ \bibnamefont {Kouwenhoven}},\ and\ \bibinfo {author}
  {\bibfnamefont {H.~S.~J.}\ \bibnamefont {van~der Zant}},\ }\bibfield  {title}
  {\bibinfo {title} {Carbon nanotubes as ultrahigh quality factor mechanical
  resonators},\ }\href {https://doi.org/10.1021/nl900612h} {\bibfield
  {journal} {\bibinfo  {journal} {Nano Letters}\ }\textbf {\bibinfo {volume}
  {9}},\ \bibinfo {pages} {2547} (\bibinfo {year} {2009})}\BibitemShut
  {NoStop}%
\bibitem [{\citenamefont {Mohanty}\ \emph {et~al.}(2002)\citenamefont
  {Mohanty}, \citenamefont {Harrington}, \citenamefont {Ekinci}, \citenamefont
  {Yang}, \citenamefont {Murphy},\ and\ \citenamefont {Roukes}}]{TLS3}%
  \BibitemOpen
  \bibfield  {author} {\bibinfo {author} {\bibfnamefont {P.}~\bibnamefont
  {Mohanty}}, \bibinfo {author} {\bibfnamefont {D.~A.}\ \bibnamefont
  {Harrington}}, \bibinfo {author} {\bibfnamefont {K.~L.}\ \bibnamefont
  {Ekinci}}, \bibinfo {author} {\bibfnamefont {Y.~T.}\ \bibnamefont {Yang}},
  \bibinfo {author} {\bibfnamefont {M.~J.}\ \bibnamefont {Murphy}},\ and\
  \bibinfo {author} {\bibfnamefont {M.~L.}\ \bibnamefont {Roukes}},\ }\bibfield
   {title} {\bibinfo {title} {Intrinsic dissipation in high-frequency
  micromechanical resonators},\ }\href
  {https://doi.org/10.1103/PhysRevB.66.085416} {\bibfield  {journal} {\bibinfo
  {journal} {Phys. Rev. B}\ }\textbf {\bibinfo {volume} {66}},\ \bibinfo
  {pages} {085416} (\bibinfo {year} {2002})}\BibitemShut {NoStop}%
\bibitem [{\citenamefont {Tiersten}(1976)}]{Tiersten1976}%
  \BibitemOpen
  \bibfield  {author} {\bibinfo {author} {\bibfnamefont {H.~F.}\ \bibnamefont
  {Tiersten}},\ }\bibfield  {title} {\bibinfo {title} {Analysis of nonlinear
  resonance in thickness-shear and tranned - anarav resonators},\ }\href
  {https://doi.org/10.1121/1.380946} {\bibfield  {journal} {\bibinfo  {journal}
  {Journal of the Acoustical Society of America}\ }\textbf {\bibinfo {volume}
  {59}},\ \bibinfo {pages} {866} (\bibinfo {year} {1976})}\BibitemShut
  {NoStop}%
\bibitem [{\citenamefont {Tiersten}\ and\ \citenamefont
  {Stevens}(1986)}]{Tiersten1986}%
  \BibitemOpen
  \bibfield  {author} {\bibinfo {author} {\bibfnamefont {H.~F.}\ \bibnamefont
  {Tiersten}}\ and\ \bibinfo {author} {\bibfnamefont {D.~S.}\ \bibnamefont
  {Stevens}},\ }\bibfield  {title} {\bibinfo {title} {{An analysis of nonlinear
  resonance in contoured-quartz crystal resonators}},\ }\href
  {https://doi.org/10.1121/1.393802} {\bibfield  {journal} {\bibinfo  {journal}
  {Journal of the Acoustical Society of America}\ }\textbf {\bibinfo {volume}
  {80}},\ \bibinfo {pages} {1122} (\bibinfo {year} {1986})}\BibitemShut
  {NoStop}%
\bibitem [{\citenamefont {Gagnepain}\ and\ \citenamefont
  {Besson}(2012)}]{nonlin}%
  \BibitemOpen
  \bibfield  {author} {\bibinfo {author} {\bibfnamefont {J.}~\bibnamefont
  {Gagnepain}}\ and\ \bibinfo {author} {\bibfnamefont {R.}~\bibnamefont
  {Besson}},\ }\bibinfo {title} {Physical acoustics}\ (\bibinfo  {publisher}
  {Elsevier},\ \bibinfo {year} {2012})\ Chap.\ \bibinfo {chapter} {Nonlinear
  Effects in Piezoelectric Quartz Crystals}, pp.\ \bibinfo {pages}
  {245--288}\BibitemShut {NoStop}%
\bibitem [{\citenamefont {Goryachev}\ \emph
  {et~al.}(2013{\natexlab{c}})\citenamefont {Goryachev}, \citenamefont {Farr},
  \citenamefont {Ivanov},\ and\ \citenamefont {Tobar}}]{quartzJAP}%
  \BibitemOpen
  \bibfield  {author} {\bibinfo {author} {\bibfnamefont {M.}~\bibnamefont
  {Goryachev}}, \bibinfo {author} {\bibfnamefont {W.}~\bibnamefont {Farr}},
  \bibinfo {author} {\bibfnamefont {E.}~\bibnamefont {Ivanov}},\ and\ \bibinfo
  {author} {\bibfnamefont {M.}~\bibnamefont {Tobar}},\ }\bibfield  {title}
  {\bibinfo {title} {Anomalously strong nonlinearity of unswept quartz acoustic
  cavities at liquid helium temperatures},\ }\href@noop {} {\bibfield
  {journal} {\bibinfo  {journal} {Journal of Applied Physics}\ }\textbf
  {\bibinfo {volume} {114}},\ \bibinfo {pages} {094506} (\bibinfo {year}
  {2013}{\natexlab{c}})}\BibitemShut {NoStop}%
\bibitem [{\citenamefont {Fillinger}\ \emph {et~al.}(2006)\citenamefont
  {Fillinger}, \citenamefont {Zaitsev}, \citenamefont {Gusev},\ and\
  \citenamefont {Castagnède}}]{Fillinger2006}%
  \BibitemOpen
  \bibfield  {author} {\bibinfo {author} {\bibfnamefont {L.}~\bibnamefont
  {Fillinger}}, \bibinfo {author} {\bibfnamefont {V.}~\bibnamefont {Zaitsev}},
  \bibinfo {author} {\bibfnamefont {V.}~\bibnamefont {Gusev}},\ and\ \bibinfo
  {author} {\bibfnamefont {B.}~\bibnamefont {Castagnède}},\ }\bibfield
  {title} {\bibinfo {title} {Wave self-modulation in an acoustic resonator due
  to self-induced transparency},\ }\href
  {https://doi.org/10.1209/epl/i2006-10267-5} {\bibfield  {journal} {\bibinfo
  {journal} {Europhysics Letters}\ }\textbf {\bibinfo {volume} {76}},\ \bibinfo
  {pages} {229} (\bibinfo {year} {2006})}\BibitemShut {NoStop}%
\bibitem [{\citenamefont {Fillinger}\ \emph {et~al.}(2008)\citenamefont
  {Fillinger}, \citenamefont {Zaitsev}, \citenamefont {Gusev},\ and\
  \citenamefont {Castagnède}}]{FILLINGER2008}%
  \BibitemOpen
  \bibfield  {author} {\bibinfo {author} {\bibfnamefont {L.}~\bibnamefont
  {Fillinger}}, \bibinfo {author} {\bibfnamefont {V.}~\bibnamefont {Zaitsev}},
  \bibinfo {author} {\bibfnamefont {V.}~\bibnamefont {Gusev}},\ and\ \bibinfo
  {author} {\bibfnamefont {B.}~\bibnamefont {Castagnède}},\ }\bibfield
  {title} {\bibinfo {title} {Self-modulation of acoustic waves in resonant
  bars},\ }\href {https://doi.org/https://doi.org/10.1016/j.jsv.2008.02.034}
  {\bibfield  {journal} {\bibinfo  {journal} {Journal of Sound and Vibration}\
  }\textbf {\bibinfo {volume} {318}},\ \bibinfo {pages} {527} (\bibinfo {year}
  {2008})}\BibitemShut {NoStop}%
\bibitem [{\citenamefont {Goryachev}\ \emph {et~al.}(2011)\citenamefont
  {Goryachev}, \citenamefont {Galliou}, \citenamefont {Imbaud}, \citenamefont
  {Bourquin}, \citenamefont {Dulmet},\ and\ \citenamefont
  {Abbe}}]{goryachev:hal-00670126}%
  \BibitemOpen
  \bibfield  {author} {\bibinfo {author} {\bibfnamefont {M.}~\bibnamefont
  {Goryachev}}, \bibinfo {author} {\bibfnamefont {S.}~\bibnamefont {Galliou}},
  \bibinfo {author} {\bibfnamefont {J.}~\bibnamefont {Imbaud}}, \bibinfo
  {author} {\bibfnamefont {R.}~\bibnamefont {Bourquin}}, \bibinfo {author}
  {\bibfnamefont {B.}~\bibnamefont {Dulmet}},\ and\ \bibinfo {author}
  {\bibfnamefont {P.}~\bibnamefont {Abbe}},\ }\bibfield  {title} {\bibinfo
  {title} {{Recent Investigations on BAW Resonators at Cryogenic
  Temperatures}},\ }in\ \href {https://doi.org/10.1109/FCS.2011.5977293} {\emph
  {\bibinfo {booktitle} {{2011 JOINT CONFERENCE OF THE IEEE INTERNATIONAL
  FREQUENCY CONTROL SYMPOSIUM/EUROPEAN FREQUENCY AND TIME FORUM
  PROCEEDINGS}}}}\ (\bibinfo {address} {San Francisco, United States},\
  \bibinfo {year} {2011})\ pp.\ \bibinfo {pages} {819--824}\BibitemShut
  {NoStop}%
\end{thebibliography}%
\end{document}